\documentclass[seceq]{ptptex}

\usepackage{graphicx}

\title{
Efficient Implementations of Molecular Dynamics Simulations for Lennard-Jones Systems
}

\author{
Hiroshi \textsc{WATANABE}$^{1,2}$$\footnote{E-mail: hwatanabe@issp.u-tokyo.ac.jp}$, %
Masaru \textsc{SUZUKI}$^3$, and %
Nobuyasu \textsc{ITO}$^4$%
}

\inst{
$^1$ Institute for Solid State Physics, The University of Tokyo, Chiba 277-8581, Japan\\
$^2$ Information Technology Center, The University of Tokyo, Tokyo 113-8558, Japan \\
$^3$Department of Applied Quantum Physics and Nuclear Engineering, Kyushu University, Fukuoka 819-0395, Japan\\
$^4$Department of Applied Physics, School of Engineering, The University of Tokyo, Tokyo 113-8656, Japan\\
}

\abst{
Efficient implementations of the classical molecular dynamics (MD) method for
Lennard-Jones particle systems are considered.
Not only general algorithms but also techniques that are efficient for some specific CPU
architectures are also explained.
A simple spatial-decomposition-based strategy is adopted for parallelization.
By utilizing the developed code, benchmark simulations are performed on a
HITACHI SR16000/J2 system consisting of IBM POWER6 processors which are 4.7 GHz at the National Institute for Fusion Science (NIFS) and 
an SGI Altix ICE 8400EX system consisting of Intel Xeon processors which are 2.93 GHz at the Institute for Solid State Physics (ISSP), the University of Tokyo.
The parallelization efficiency of the largest run, consisting of 4.1 billion particles with 8192 MPI processes, is about 73\%
relative to that of the smallest run with 128 MPI processes at NIFS, and it is about 66\% relative to that of the smallest run with 4 MPI processes at ISSP.
The factors causing the parallel overhead are investigated.
It is found that fluctuations of the execution time of each process degrade the parallel efficiency.
These fluctuations may be due to the interference of the operating system, which is known as OS Jitter.
}

\newcommand{\TO}{\textbf{to}~}

\usepackage{algorithm,algorithmic}

\begin{document}

\maketitle

\section{Introduction}

The classical molecular dynamics (MD) simulation was first performed by Alder and Wainwright~\cite{Alder1957}.
The original MD was used to investigate the hard-particle system, and an event-driven algorithm was adopted for the time evolution,
which was soon followed by a time-step-driven algorithm~\cite{Rahman1964}.
Owing to the recent increase in computational power, MD is now a powerful tool for studying
not only the equilibrium properties but also the nonequilibrium transportation phenomena of molecular systems~\cite{Allen, Frenkel, Rapaport,HooverBook}
as well as biomolecular systems~\cite{NAMD, GROMACS}.
Electronic degrees of freedom can also be considered by \textit{ab initio} methods on the basis of quantum mechanics~\cite{Car1985}.
The increase in computational power also allows us not only to use more realistic and complex interactions
but also to treat a larger number of particles.
The recent increase in computational power has mainly been achieved by increasing the number of processing cores.
This paradigm is called massively parallel processing (MPP), and
most recent high-end computers have been built on the basis of this paradigm~\cite{Top500}.
The number of cores is typically from several thousand to several hundred thousand.
In an MPP system, a single task is performed by a huge number of microprocessors, which operate simultaneously and
communicate with each other as needed.
Therefore, parallelization is now unavoidable in order to utilize
the computational power of such machines effectively.
While parallelization itself is of course important,
the tunings for a single core have become more important for huge-scale and long-time simulations,
since the overall performance of MD mainly depends on the performance on the single cores.
In particular, optimizing memory access is important.
The latency and bandwidth of memory access are sometimes slow compared with the execution speed of processors,
and the data supply often cannot keep up with the requests by the processors.
Therefore, data must be suitably arranged so that necessary data is located near a processor.
Tuning for specific architectures is also important to achieve high performance.
Since design concepts differ from one processor to another,
it is necessary to prepare codes for each architecture, at least for the hot spot of the simulation
where most time is spent during the execution, which is usually force calculation in MD.

The Next-Generation Supercomputer project is currently being carried out by RIKEN~\cite{RIKEN}.
The system is designed on the basis of the MPP paradigm and will use scalar CPUs
with 128 GFlops, will consist of over 80 000 nodes, 
and is expected to achieve a total computational power of 10 petaflops.
This supercomputer is now under development and will be ready in 2012.
Therefore, we must prepare parallelized codes that can utilize the full performance of this system by then.
The parallelization of MD has been discussed over several decades with the aims of treating larger systems
and performing simulations for longer timescales~\cite{Plimpton1995}, and
a huge MD simulation involving one trillion particles has been performed
on BlueGene/L, which consists of 212 992 processors~\cite{Germann2008}.
Despite this success, it will not easy to obtain a satisfactory performance on the Next-Generation Supercomputer
since the properties of the systems are considerably different from those of BlueGene.
BlueGene has relatively slow processors with a 700 MHz clock speed for BlueGene/L,
and consequently, it achieves a good balance between the processor speed and the memory bandwidth.
In comparison, the Next-Generation Supercomputer will have a relatively high peak performance
of 128 GFlops per CPU (16 GFlops for each core and each CPU consists of eight cores).
Therefore, the memory bandwidth and latency may be a significant bottleneck.

The purpose of the present article is to describe efficient algorithms and
their implementation for the MD method,
particularly focusing on the memory efficiency and the parallel overheads.
We also provide some tuning techniques for a couple of the specific architectures.
While the Lennard-Jones potential is considered throughout the manuscript,
the techniques described here are generally applicable
to other simulations with more complicated potentials.
This manuscript is organized as follows.
In Sec.~\ref{sec_pairlist}, algorithms for finding interacting particle pairs are described.
Some further optimization techniques whose efficiency depends on architecture of the computer
are explained in Sec.~\ref{sec_further_optimization}.
Parallelization schemes and the results of benchmark simulations are described in Sec.~\ref{sec_parallelization}.
The factors causing the parallel overhead for the developed MD code are discussed in Sec.~\ref{sec_parallel_overhead},
and a summary and a discussion of further issues are given in Sec.~\ref{sec_summary}.

\begin{figure}[tb]
\begin{center}
\includegraphics[width=0.9\linewidth]{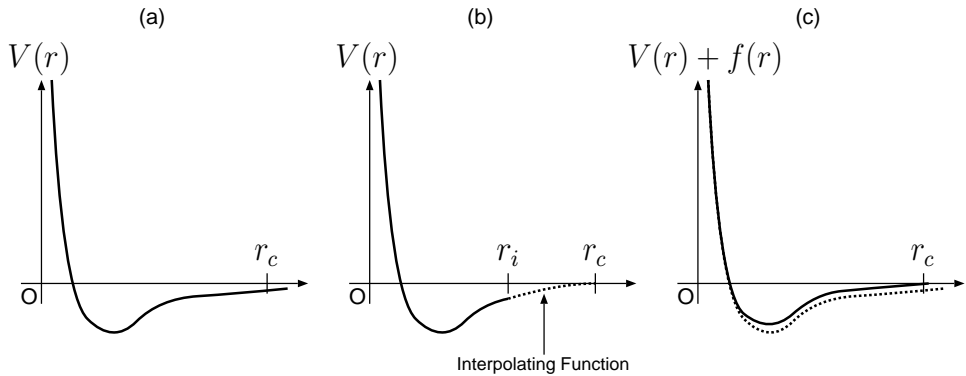}
\end{center}
\caption{
Truncation of the Lennard-Jones potential function.
(a) Original form. The interaction range extends to infinity.
(b) Truncation by interpolation. An interpolating function is introduced for the region from the interpolation point $r_\mathrm{i}$
to the truncation point $r_\mathrm{c}$.
(c) Truncation by adding extra terms. The additional function $f(r)$ is chosen so that
both the value and the derivatives of $V(r) + f(r)$ are zero at the truncation point.
}
\label{fig_truncation}
\end{figure}

\section{Pair List Construction}
\label{sec_pairlist}

\subsection{Truncation of Potential Function} \label{sec_potential}

The common expression for the Lennard-Jones potential is
\begin{equation}
V(r) = 
\displaystyle 4 \varepsilon \left[
\left( \frac{\sigma}{r} \right)^{12} -
\left( \frac{\sigma}{r} \right)^{6}
\right], \label{eq_lj_original}
\end{equation}
for the particle distance $r$, well depth $\varepsilon$, and atomic diameter $\sigma$.
In the following, we will use the physical quantities reduced by $\sigma$, $\varepsilon$, and $k_\mathrm{B}$,
for example., the length scale is measured in the unit of $\sigma$, and so forth.
The first term of the right-hand side of Eq.~(\ref{eq_lj_original}) describes the repulsive force, and the other term describes the attractive force.
This potential form has been widely used for a standard particle model involving phase transitions since
this system exhibits three phases: solid, liquid, and gas.
While the original form of the Lennard-Jones potential spans infinite range, 
it is wasteful to consider an interaction between two particles at a long distance since the potential decays rapidly as the distance increases.
To reduce computation time, truncated potentials are usually used instead of the original potential\cite{Allen}.
There are several ways of introducing truncation. One way is to use cubic spline interpolation for the region
from some distance to the cutoff length\cite{Holian1991, Beazley1994} as shown in Fig.~\ref{fig_truncation}~(b).
In this scheme, there are two special distances, the interpolating distance $r_\mathrm{i}$
and the truncation distance $r_\mathrm{c}$. 
The potential is expressed by a piecewise-defined function that 
switches from $V(r)$ given in  Eq.~(\ref{eq_lj_original})
to an interpolating function $V_\mathrm{i}(r)$ at $r = r_\mathrm{i}$.
The interpolation function is chosen so that values and the derivatives of the potential
are continuous at the interpolation and the truncation points.
These conditions require four adjustable parameters, and therefore,
a third-order polynomial is usually chosen as the interpolating function.
While this interpolation scheme was formerly popular, it is now outdated
since it involves a conditional branch, which is sometimes expensive for current CPU architectures.

Nowadays, truncation is usually introduced by adding some extra terms to the potential.
At the truncation point, both the potential and the force should become continuous,
and therefore, at least two additional terms are necessary.
One such potential is given by~\cite{Spotswood1973}
\begin{equation}
V(r) = 
\displaystyle 4 \varepsilon \left[
\left( \frac{\sigma}{r} \right)^{12} -
\left( \frac{\sigma}{r} \right)^{6} +
c_2 \left( \frac{r}{\sigma} \right)^2+ c_0 \right], \label{eq_lj_cutoff}
\end{equation}
with two additional coefficients $c_2$ and $c_0$, which are determined so that
$V(r_\mathrm{c}) = V'(r_\mathrm{c}) = 0$ for the cutoff length $r_\mathrm{c}$.
Here, we use a quadratic term instead of a linear term such as  $c_1 r + c_0$, 
since the latter involves a square-root operation in the force calculation, which is sometimes expensive.
While the computational cost of the force computation is more expensive than that for the interpolating method,
this method is usually faster than the interpolating method since conditional branches can be hazards
in the pipeline that obstruct the instruction stream.
If accuracy is less crucial, one can truncate the potential by adding one constant,
\textit{i.e.}, $c_2 = 0$ and $c_0$ is determined by $V(r_\mathrm{c}) = 0$.
In this case, the calculation of force is identical to that using the original potential and the modification only appears in the calculation of the potential energy.
Note that this scheme sometimes involves some problems regarding the conservation of energy since the force is not continuous at the truncation point.
The truncation changes the phase diagram.
One should check the phase diagram and compute phase boundaries for each truncated potential before performing production runs.
In particular, the gas-liquid coexistence phase becomes drastically narrower as the cutoff length decreases.
To investigate phenomena involving the gas-liquid phase transition,
the cutoff length should be made longer than $3.0 \sigma$.

\subsection{Grid Search Method}
\label{sec_grid_search}

\begin{figure}[tb]
\begin{center}
\includegraphics[width=0.45\linewidth]{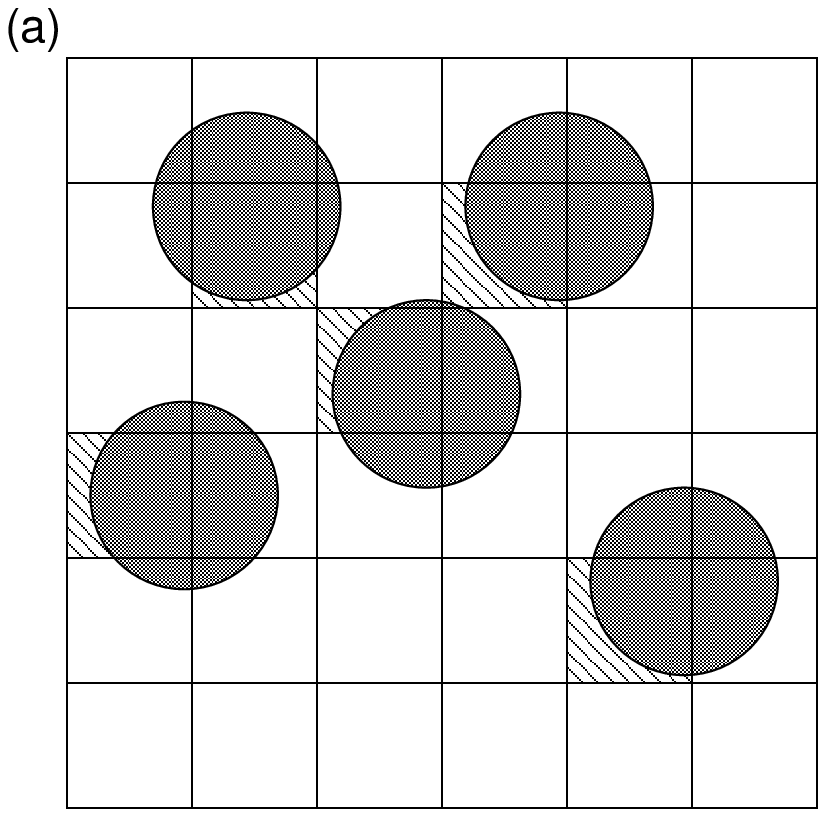}
\includegraphics[width=0.45\linewidth]{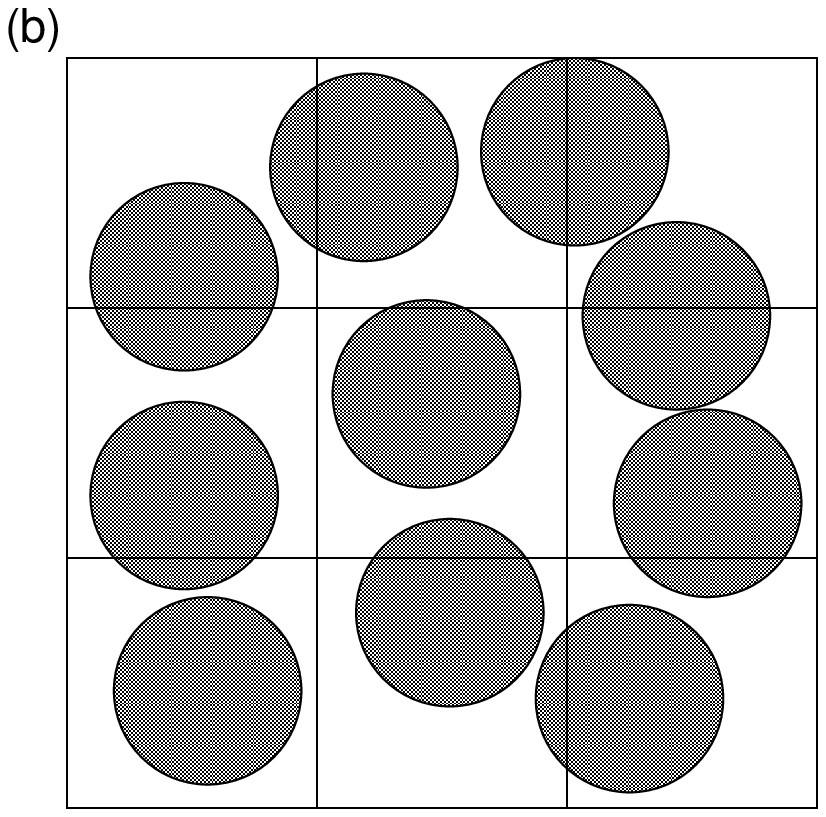}
\end{center}
\caption{
Searching for interacting pairs in a grid.
(a) Exclusive grid. The size of each cell is determined so that only one particle can be placed in each cell.
(b) Inexclusive grid. The interaction length determines the size of cells.
More than one particle is allowed to occupy one cell simultaneously.
}
\label{fig_grid}
\end{figure}

To perform the time evolution of a system,
we first have to find particle pairs such that the distance between the particles is less than the cutoff length.
Since the trivial computation leads to $O(N^2)$ computation, where $N$ is the number of particles,
we adopt a grid algorithm, which is also called the linked-list method~\cite{Quentrec1975, Hockney1981}, to reduce the complexity of the computation to $O(N)$.
The algorithm finds interacting pairs with in a grid by the following three steps:
(i) divide a system into small cells.
(ii) register the indices of particles on cells to which they belong, and
(iii) search for interacting particle pairs using the registered information.
There are two types of grid, one is exclusive and the other is inexclusive (see Fig.~\ref{fig_grid}).
An exclusive grid allows only one particle to occupy a cell~\cite{Form1993}, and the inexclusive grid allows
more than one particle to occupy one cell simultaneously~\cite{Allen,Knuth1973,Grest1989,Beazley1994}.
Generally, an exclusive grid is efficient for a system with short-range interactions
such as hard particles, and an inexclusive grid is suitable for a system with relatively long interactions
such as Lennard-Jones particles. However, the efficiency strongly depends not only on
the physical properties such as density, polydispersity, and interaction length but also on the hardware architecture such as the
cache size and memory bandwidth. Therefore, it is difficult to determine which is better before implementation.
We have found that an inexclusive grid is more efficient than an exclusive grid for the region
in which a Lennard-Jones system involves a gas-liquid phase transition.
Therefore, we adopt an inexclusive grid in the following.
Note that the length of each cell should be larger than the search length $r_\mathrm{s}$, which is longer than the cutoff length as described later,
so that there are no interactions occurring through three or more cells, in other words, 
interacting particle pairs are always in the same cell or adjacent cells.

\begin{figure}[tb]
\begin{center}
\includegraphics[width=0.52\linewidth]{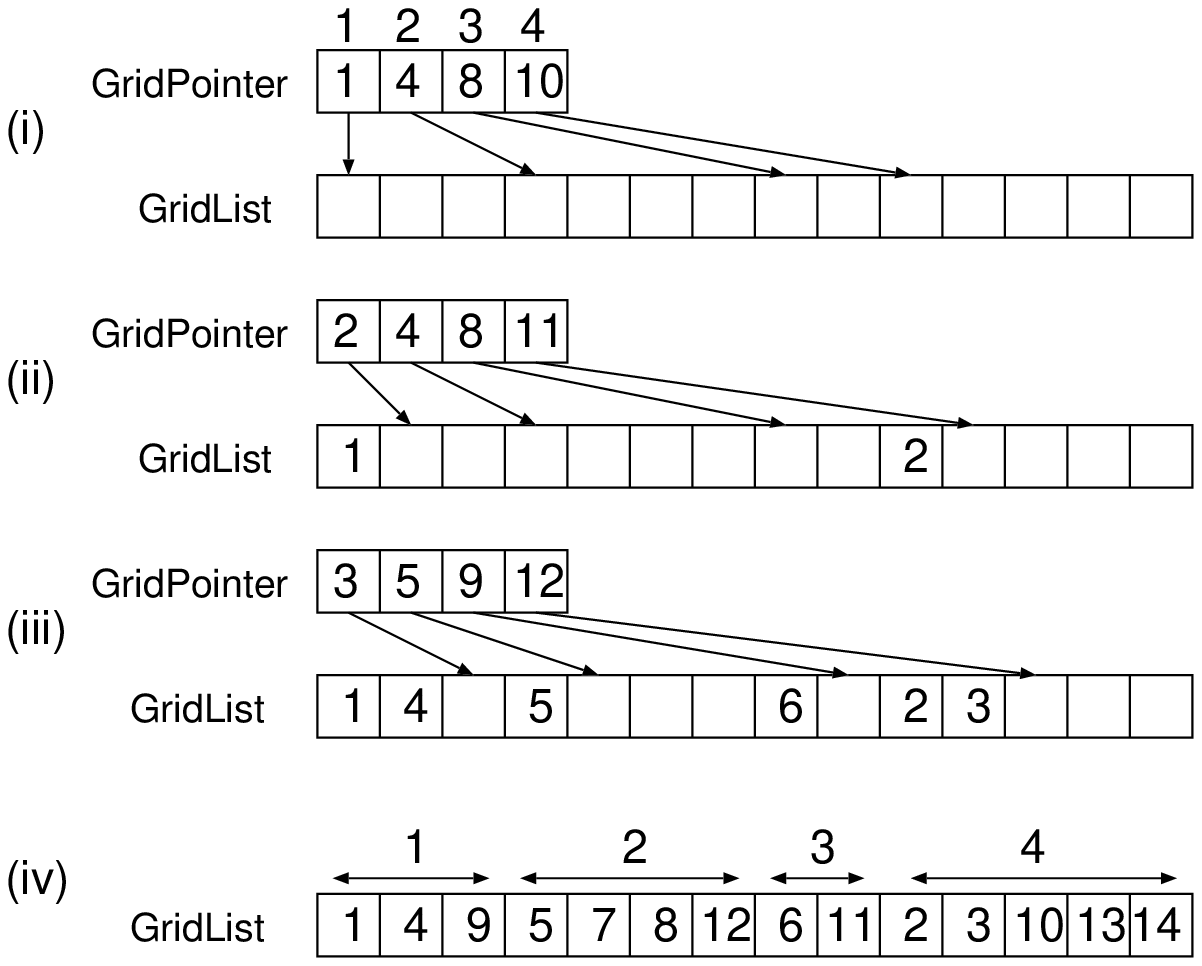}
\includegraphics[width=0.44\linewidth]{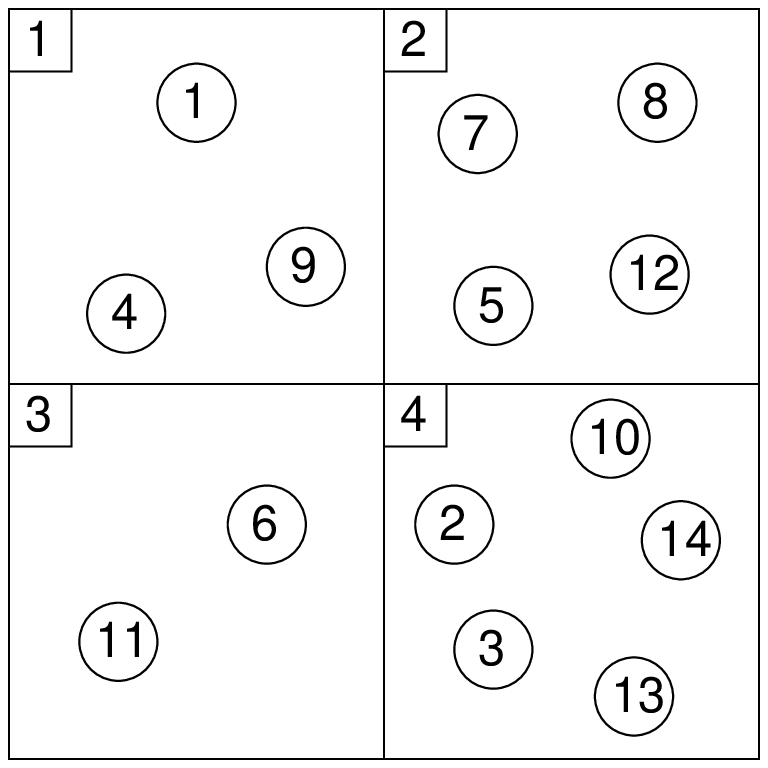}
\end{center}
\caption{
How to construct GridList, which is a linear array of size $N$ storing grid information.
Suppose there are four cells and $14$ particles in a system ($N=14$).
The left figure shows the stored data in the array and 
the right figure shows the configuration of the particles in the system.
The numbers with arrows denote pointers indicating where the indices of particles should be stored (GridPointer).
The particles have indices from $1$ to $14$ and
the cells are labeled from $1$ to $4$. 
First, prepare an array of size $N$, and label particles with the serial number of the cells to which they belong.
(i) Count the number of particles in each cell, store the number as GridParticleNumber, and set a pointer for each cell at the position
where the index of the first particle of each grid should be stored when the particles are sorted by their labels.
(ii) Place the index of a particle at the position where the pointer corresponding to its label points, and move the pointer to the right.
(iii) Repeat this procedure for all particles. 
The status of the GridList just after the particle number 6 is located into the GridList.
(iv) The completed array after above (i) -- (iii) procedures. The particles are sorted in GridList
in order of the grid labels.
}
\label{fig_gridlist}
\end{figure}

A simple way to express the grid information is to use multidimensional arrays.
Two types of array are necessary, one is for the number of particles in each cell, and the other is for the indices of the particles in each cell.
Suppose the total number of cells is $n_g = n_{gx} \times n_{gy} \times n_{gz}$, then the arrays can be expressed by
\begin{equation}
\begin{array}{ll}
\mathrm{integer}: & \mathrm{GridParticleNumber}[n_{gx}][n_{gy}][n_{gz}] \\
\mathrm{integer}: & \mathrm{GridIndex}[n_{gx}][n_{gy}][n_{gz}][g_\mathrm{max}],
\end{array}
\end{equation}
where $g_\mathrm{max}$ is the capacity of one cell.
The number of particles in a cell at $(x, y, z)$ is stored as  $\mathrm{GridParticleNumber}[x][y][z]$
and the index of the $i$-th particle at the cell is sotred as $\mathrm{GridIndex}[x][y][z][i]$.
Although this scheme is simple, the required memory can be dozens of times larger than the total number of particles in the system,
and therefore, the indices of the particles are stored sparsely, which causes a decrease in cache efficiency.
To improve the efficiency of memory usage, the indices of particles should be stored sequentially 
in an array. This method is proposed by Kadau \textit{et al.} and implemented in SPaSM~\cite{Kadau2006}.
An algorithm to construct grid information using an array whose size is the total number of particles $N$ is as follows.
\begin{algorithmic}[1]
\STATE Divide a system into small cells, and assign a serial number to each cell.
\STATE Label particles with the serial numbers of the cells that they belong to.
\STATE Count the number of particles in each cell.
\STATE Sort the particles by their labels.
\end{algorithmic}
The details of the implementation are shown in Fig.~\ref{fig_gridlist}.
Suppose that $i$-cell denotes the cell labeled with $i$ and the grid information is stored in the linear array GridList.
Then GridParticleNumber[$i$] denotes the number of particles in $i$-cell and
GridPointer[$i$] denotes the position of which the first particle of $i$-cell should be stored in GridList.
The complexity of this sorting process is $O(N)$ instead of $O(N \log N)$ for usual Quicksort
since it does not perform a complete sorting.
The memory usage is effective since the indices of particles in each grid are stored contiguously in the array.

\begin{figure}[tb]
\begin{center}
\includegraphics[width=0.5\linewidth]{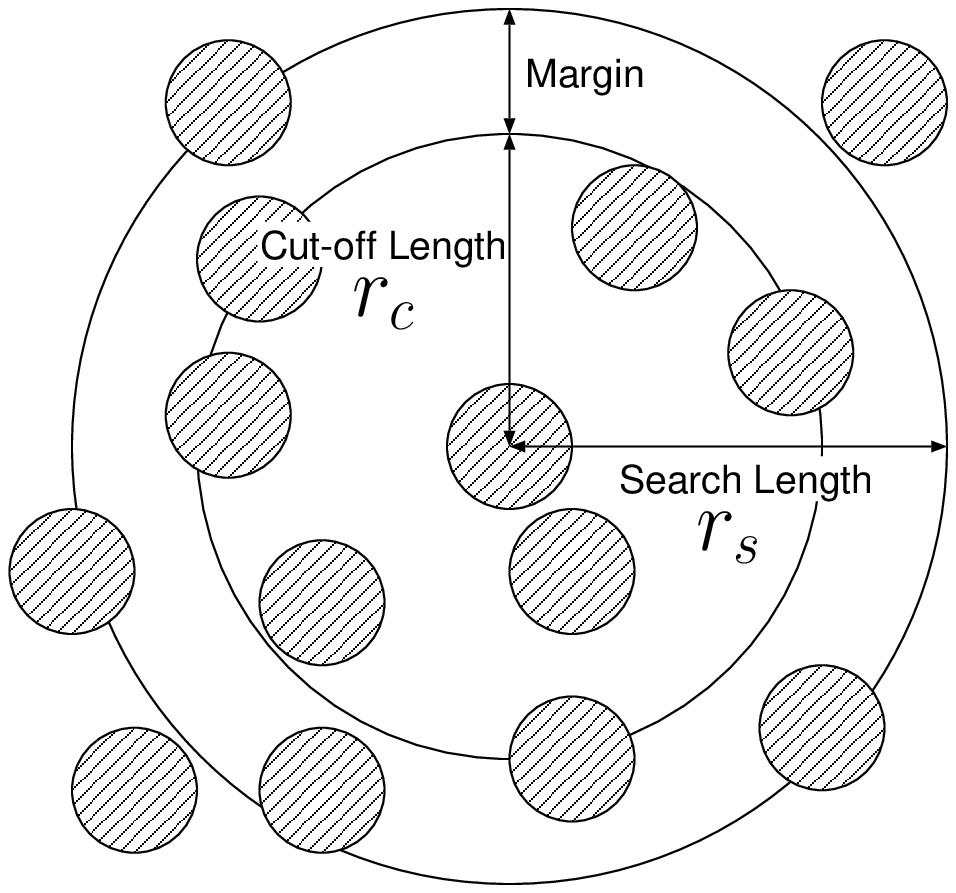}
\end{center}
\caption{
Cutoff length and search length.
The search length $r_\mathrm{s}$ is chosen so that $r_\mathrm{s} > r_\mathrm{c}$ for the cutoff length $r_\mathrm{c}$.
A pair list can be reused unless a particle located outside of search length across half of the margin $r_\mathrm{s} - r_\mathrm{c}$.
}
\label{fig_search}
\end{figure}

\begin{algorithm}[tb]
\caption{Finding the interacting particle pairs}
\label{alg_findpair}
\begin{algorithmic}[1]
\FORALL{$i$-cell}
\STATE $s_i \leftarrow$  GridPointer[$i$] 
\STATE $e_i \leftarrow$  $s_i +$ GridParticleNumber[$i$] -1
\STATE $L_\mathrm{c} \leftarrow $GridList[$s_i .. e_i$] 
\FORALL{$j$ such that $j$-cell is the neighbor of $i$-cell}
\STATE $s_j \leftarrow$  GridPointer[$j$] 
\STATE $e_j \leftarrow s_j +$  GridParticleNumber[$j$] -1
\STATE Append  GridList[$s_j .. e_j$]  to  $L_\mathrm{c}$ 
\ENDFOR
\FOR{$k=1$ \TO GridParticleNumber[$i$]}
\FOR{$l=k+1$ \TO $|L_\mathrm{c}|$ }
\STATE$m \leftarrow L_\mathrm{c}$[$k$]
\STATE$n \leftarrow L_\mathrm{c}$[$l$]
\IF  {$m$- and $n$-particles interact each other}
\STATE Register a pair $(m,n)$ to PairList
\ENDIF
\ENDFOR
\ENDFOR
\ENDFOR
\end{algorithmic}
\end{algorithm}

\subsection{Bookkeeping Method}
\label{sec_dutc}

After constructing GridList, we have to find interacting particle pairs using the grid information.
Suppose the particle pairs whose distance is less than the cutoff length are stored in PairList.
PairList is an array of pairs which hold two indicies of interacting particles.
All particles are registered in cells and labeled by the serial numbers of the cells.
There are two cases, interacting particles share or do not share the same label,
which correspond to inner-cell interaction and cell-crossing interaction.
Let GridList[$s .. e $] denote a subarray of GridList whose range is from $s$ to $e$ with
the data of the pair list stored in PairList.
The pseudocodes used to find interacting pairs are shown in Algorithm~\ref{alg_findpair}.
Only 13 of the 26 neighbors should be checked for each cell because of Newton's third law.

As the number of particles increases, the cost of constructing the pair list increases drastically
since it involves frequent access to memory. 
While it depends on the details of the system,
constructing a pair list is sometimes dozens of times more expensive than the force calculation.
In order to reduce the cost of constructing a pair list, a well-known bookkeeping method was proposed~\cite{Verlet1967}.
The main idea of the bookkeeping method is to reuse the same pair list for several time steps
by registering pairs within a search length $r_\mathrm{s}$ which is set at longer than the cutoff (interaction) length $r_\mathrm{c}$ (see Fig.~\ref{fig_search}).
The margin $r_\mathrm{s} - r_\mathrm{c}$ determines the lifetime of a pair list.
While a longer margin gives a longer lifetime, the length of the pair list also increases,
and consequently, the computational time for the force calculation becomes longer.
Therefore, there is an
optimal length for the margin that depends on the details of the system, such as its density and temperature.
Several steps after the construction of the pair list, the list can become invalid, \textit{i.e.},
some particle pair that is not registered in the list may be at a distance shorter than the interaction length $r_\mathrm{c}$.
Therefore, we have to calculate the expiration time of the list and can check the validity of the pair list for each step as the following simple method.
The expiration time $t_\mathrm{e}$ of a pair list constructed at time $t$ is given by
\begin{equation}
t_\mathrm{e} = \frac{r_\mathrm{s} - r_\mathrm{c}}{2 v_\mathrm{max}} +t,
\end{equation}
where $v_\mathrm{max}$ is the absolute value of the velocity of the fastest particle in the system.
The factor $2$ in the denominator corresponds to the condition that two particles undergo a head-on collision.
It is necessary to update the expiration time $t_\mathrm{e}$ when the maximum velocity changes to
\begin{equation}
t_\mathrm{e} \leftarrow (t_\mathrm{e} - t) \frac{v_\mathrm{old}}{v_\mathrm{new}} +t,
\end{equation}
where $v_\mathrm{old}$ and $v_\mathrm{new}$ denote the previous and current maximum speeds, respectively.
Obviously, the expiration time will be brought forward in the case of a faster particle, and vice versa.
This technique for calculating the expiration time is called the
dynamical upper time cutoff (DUTC) method and was proposed by Isobe~\cite{Isobe1999}.

\begin{figure}[tb]
\begin{center}
\includegraphics[width=0.9\linewidth]{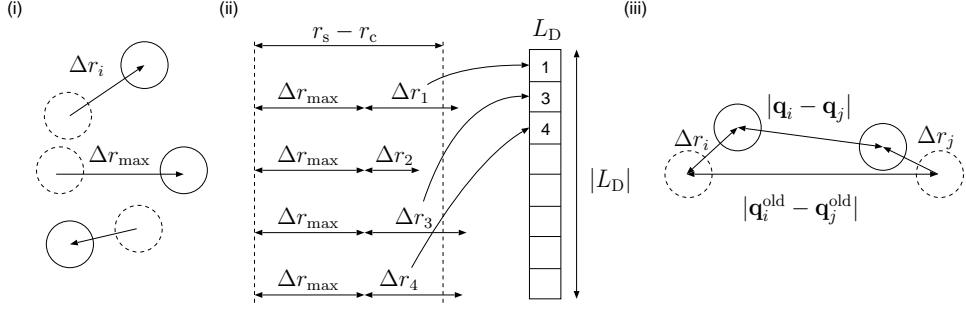}
\end{center}
\caption{
The large displacement check (LDC) method. 
(i) Calculate the displacements $\Delta r_i$ for the particle $i$ and find the largest displacement $\Delta_\mathrm{max}$.
(ii) Register a index of a particle $i$ such that $\Delta r_i > r_\mathrm{s} - r_\mathrm{c} - \Delta r_\mathrm{max}$ to a list $L_\mathrm{D}$ whose size is denoted by $|L_\mathrm{D}|$.
(iii) Old and current position of particles $i$ and $j$. The dashed circle denote the old position
which was registered in the previous consturction of the pair list.
The solid circles denote the current positions of particles.
The expiration of the pair list is determined by the old and the current distance of the particles.
}
\label{fig_ldc}
\end{figure}

Although the validity check by considering the maximum velocity is simple,
the expiration time  $t_\mathrm{e}$ determined by the velocities of the particles
 is shorter than the \textit{actual} expiration time of the pair list, since the pair list is valid 
as long as the particles migrate short distances.
Therefore, we propose a method which extends the lifetime of a pair list by utilizing the displacements of the particles.
The strict condition for the expiration of a pair list is that
there exists a particle pair $(i,j)$ such that
\begin{equation}
|\textbf{q}_i^{\mathrm{old}} - \textbf{q}_j^{\mathrm{old}}| > r_\mathrm{s} \qquad \mbox{and}  \qquad
|\textbf{q}_i - \textbf{q}_j| < r_\mathrm{c}, \label{eq_strict_condition}
\end{equation}
where $\textbf{q}_i^{\mathrm{old}}$ is the position of the $i$-particle when the pair list was constructed
and $\textbf{q}_i$ is the current position of the $i$-particle. 
The lifetime of a pair list can be extended if we use the strict condition for the validity check,
but the computational cost of checking this the condition is $O(N^2)$.
We therefore extend the lifetime of the pair list by using a less strict condition.
The main idea is to prepare a list of particles that have moved by large amounts and check the distance between all pairs in the list.
The algorithm is as follows:
(i) Calculate the displacements of the particles from their position when the pair list was constructed
and find the largest displacement $\Delta r_\mathrm{max}$.
(ii) If there exists a particle such that $\Delta r_i > r_\mathrm{s} - r_\mathrm{c} - \Delta r_\mathrm{max}$, then
append the particle to the list of large displacement $L_\mathrm{D}$ which size is denoted by $|L_\mathrm{D}|$.
If the list is longer than the predetermined maximum length of the list $N_\mathrm{l}$,
that is, $|L_\mathrm{D}| > N_\mathrm{l}$  ,then the pair list is invalidated.
(iii) If the old distance is longer than the search length and the current distance is shoter than the 
cutoff length, it means that there are interacting particles which are not registered in a pair list.
Therefore, the pair list is invalidated if there exists a pair $(i, j)$ in the list of large displacement such that
$ |\textbf{q}_i^{\mathrm{old}} - \textbf{q}_j^{\mathrm{old}}| > r_\mathrm{s} $ and  $|\textbf{q}_i  - \textbf{q}_j| < r_\mathrm{c}$.
We call this method the large displacement check (LDC) method. See Fig.~\ref{fig_ldc} for a schematic description of the LDC method.

Although the lifetime of the pair list increases with $N_l$,
the computational cost also increases as $O(N_l^2)$. Therefore, there is an optimal value for $N_l$.
Since the LDC method is more expensive than DUTC method,
we adopt a hybrid algorithm involving both methods, that is,
first we use the DUTC method for the validity check,
and then we use LDC method after the DUTC method decides that the pairlist is expired.
The full procedure of the validity check is shown in Algorithm~\ref{alg_validity_check}.

\begin{algorithm}
\caption{Checking the validity of a pair list}
\label{alg_validity_check}
\begin{enumerate}
\item When a pair list is constructed
\begin{algorithmic}[1]
\FORALL{$i$-particle}
\STATE $\textbf{q}_i^{\mathrm{old}} \leftarrow \textbf{q}_i $ \COMMENT{Keep the positions}
\ENDFOR
\STATE buffer\_length $\leftarrow r_\mathrm{s} - r_\mathrm{c}$
\end{algorithmic}
\item Validity check by DUTC method
\begin{algorithmic}[1]
\STATE $v_\mathrm{max} \leftarrow $ the maximum velocity of the particle
\STATE buffer\_length  $ \leftarrow$ buffer\_length$ - 2 v_\mathrm{max} \Delta t$
\IF {buffer\_length $ > 0$}
\STATE the pair list is valid.
\ELSE
\STATE Check the validity by considering displacements.
\ENDIF
\end{algorithmic}
\item Validity check by LDC method
\begin{algorithmic}[1]
\FORALL{$i$-particle}
\STATE $\Delta r_i \leftarrow |\textbf{q}_i^{\mathrm{old}} - \textbf{q}_i |$
\ENDFOR
\STATE $\Delta r_\mathrm{max} \leftarrow \max \{\Delta r_i  \}$
\STATE Prepare a list $L_\mathrm{D}$
\FORALL{$i$-particle}
\IF{$\Delta r_i  >  r_\mathrm{s} - r_\mathrm{c} - \Delta r_\mathrm{max} $}
\STATE Append $i$-particle to $L_\mathrm{D}$
\IF { $|L_\mathrm{D}| > N_l$ }
\STATE the pair list is invalidated.
\ENDIF
\ENDIF
\ENDFOR
\FORALL {$(i,j)$ in $L_\mathrm{D}$}
\IF { $ |\textbf{q}_i^{\mathrm{old}} - \textbf{q}_j^{\mathrm{old}}| > r_\mathrm{s} $ and  $|\textbf{q}_i  - \textbf{q}_j| < r_\mathrm{c}$ } 
\STATE the pair list is invalidated.
\ENDIF
\ENDFOR
\STATE the pair list is invalidated.
\end{algorithmic}
\end{enumerate}
\end{algorithm}

\subsection{Sorting of pair list}
\label{sec_sorted_pair list}

\begin{figure}
\begin{center}
\includegraphics[width=0.5\linewidth]{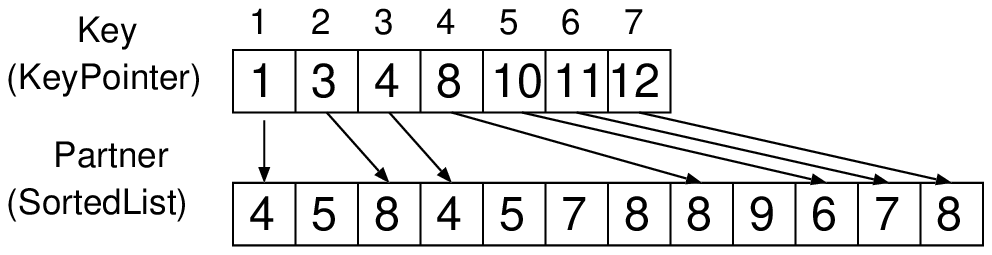}
\end{center}
\caption{
A sorted list.
A list sorted by the indices of \textit{key} particles is constructed by the following procedure.
(i) Count the number of \textit{partner} particles for each \textit{key} particle for the particle pairs are stored in the pair list.
(ii) Prepare a linear array of size  $N_\mathrm{pair}$, and set a pointer for each \textit{key} particle where the index of the first \textit{partner} particle
will be stored.
(iii) Similarly to in the grid construction~(Fig.~\ref{fig_gridlist}), place the index of a particle at the pointer of its \textit{key} particle and
move the position of the pointer to the next position.
(iv) Repeat this procedure for all pairs to construct the sorted list.
The figure shows two lists, KeyPointer and SortedList.  KeyPointer[$i$] stores the first position of the 
interacting particles in SortedList. 
}
\label{fig_sortlist}
\end{figure}

After constructing a pair list, we can calculate the forces between interacting pairs of particles.
Hereafter, we call the particle of index $i$ the $i$-particle for convenience.
Suppose the positions of particles are stored in $\textbf{q}[N]$.
A simple method for calculating the force using the pair list is shown in Algorithm~\ref{alg_calcforce}.
\begin{algorithm}
\caption{Calculating the force in a simple manner}
\label{alg_calcforce}
\begin{algorithmic}[1]
\FORALL{ pairs $(i,j)$ in PairList}
\IF{ $|\textbf{q}[i] - \textbf{q}[j]|  < r_\mathrm{c} $ }
\STATE Calculate force between $i$- and $j$-particles.
\ENDIF
\ENDFOR
\end{algorithmic}
\end{algorithm} 
This simple algorithm is, however, usually inefficient since it involves random access to the memory.
Additionally, it fetches and stores the data of both particles of each pair, which is wasteful.
Therefore, we construct a sorted list to
improve the efficiency of memory usage by calculating particles together that interact with the same particles.
For convenience, we call the lower-numbered particle in a pair the \textit{key} particle and the other the \textit{partner} particle.
The \textit{partner} particles with the same \textit{key} particle are grouped together.
The detail of the implementation is shown in Fig.~\ref{fig_sortlist}.
Suppose the data of the sorted list and the position of the first \textit{partner} particle of the $i$-particle in the list
are stored in SortedList and KeyPointer[$i$], respectively.
Then the algorithm used to calculate the force using the sorted pair list is shown in Algorithm~\ref{alg_sortedlist}.
\begin{algorithm}
\caption{Calculating the force using the sorted pair list}
\label{alg_sortedlist}
\begin{algorithmic}[1]
\FOR{$i= 1$ \TO $N-1$}  
\STATE $\textbf{q}_\mathrm{key} \leftarrow \textbf{q}[i]$
\STATE $\textbf{f}_\mathrm{key} \leftarrow 0$
\FOR{$k=$ KeyPointer[$i$] \TO KeyPointer[$i+1$]  -1 } 
\STATE $j \leftarrow $ SortedList[$k$]
\STATE $r \leftarrow |\textbf{q}_\mathrm{key} - \textbf{q}[j]| $ 
\STATE $f \leftarrow - V'(r)$
\STATE Update momenta of the $j$-particle using $f$.
\STATE Update $\textbf{f}_\mathrm{key}$ using $f$. 
\ENDFOR
\STATE Update momenta of $i$-particle using $\textbf{f}_\mathrm{key}$.
\ENDFOR
\end{algorithmic}
\end{algorithm}
The variables $\textbf{q}_\mathrm{key}$ and $\textbf{f}_\mathrm{key}$ are intended to be stored in CPU registers.
The amount of memory access decreases compared with that for Algorithm~\ref{alg_calcforce},
since the fetching and storing are performed only for the data of \textit{partner} particles in the inner loop.
The amount of memory access becomes half when the \textit{key} particles has an enough number of \textit{partner} particles.
The force acting on the \textit{key} particle is accumulated during the inner loop, and 
the momentum of the \textit{key} particle is updated at the end of the inner loop.
Note that the loop variable $i$ takes values from $1$ to $N-1$, instead of $1$ to $N$, since 
this is the loop for the \textit{key} particle, and the $N$-particle cannot be a $key$ particle.
The value of KeyPointer[$N$] should be $|\mathrm{PairList}|+1$, where $|\mathrm{PairList}|$ is the number of pairs stored in PairList.

\section{Further Optimization} 
\label{sec_further_optimization}

\subsection{Elimination of Conditional Branch}

Since we adopt the bookkeeping method, a pair list contains particle pairs whose distances are longer than the interaction length.
Therefore, it is necessary to check whether or not the distance between each pair is less than the interaction length before the force calculation (Algorithm~\ref{alg_calcforce}).
These conditional branches (so-called, "if then else" of "case" in the high-level language statement) can sometimes 
be expensive for RISC(Reduced Instruction Set Computer)-based processors such as IBM POWER6,
and it causes a significant decrease in execution speed.
To prevent the decrease in speed due to conditional branches, 
we calculate the forces without checking the distance 
and assign zero to the calculated value for the particle pairs whose distances
are longer than the interaction length.
This method is shown in Algorithm~\ref{alg_if_elimination}. 
\begin{algorithm}
\caption{Calculating the force with if-branch elimination}
\label{alg_if_elimination}
\begin{algorithmic}[1]
\FOR{$k = 1$\TO $N_\mathrm{pair}$}
\STATE $(i,j)$ $\leftarrow$ PairList[$k$]
\STATE $r$ $\leftarrow |\textbf{q}[i] - \textbf{q}[j]| $  
\STATE $f \leftarrow -V'(r)$ 
\IF{ $r \ge r_\mathrm{c}$ }
\STATE $f \leftarrow 0$
\ENDIF
\STATE Update momenta of $i$- and $j$-particles using $f$.
\ENDFOR
\end{algorithmic}
\end{algorithm}
While it appears to be wasteful, it can be faster than the original algorithm when
the penalty due to the conditional branches is more expensive than the additional cost of calculating forces.
Additionally, this loop can be performed for all pairs, which makes it easy for a compiler to optimize codes,
such as that for prefetching data. Note that, some CPU architectures prepare the instruction for
the conditional substitution.
For example, PowerPC architecture has \verb|fsel| (Floating-Point Select) instruction
which syntax is \verb|fsel FRT, FRA, FTC, FRB| where \verb|FRT, FRA, FTC|, and \verb|FRB| are registers.
If \verb|FRA| $>0$ then $\verb|FRT| = \verb|FRB|$, otherwise $\verb|FRT| = \verb|FRC|$.
We found that this optimization provides us with more than double the execution speed on IBM POWER6
in the benchmark simulation whose details are described in Sec.~\ref{sec_benchmarkresults}.

\subsection{Reduction of Divisions}

The explicit expression of Algorithm~\ref{alg_calcforce} is shown in Algorithm~\ref{alg_calcforce_explicit},
where $c_2$ is the coefficient for the truncation defined in Eq.~(\ref{eq_lj_cutoff}) and
$dt$ is the time step. 
For simplicity, the diamiter of the particle $\sigma$ is set to unity and 
only the force calculations are shown.
\begin{algorithm}
\caption{Explicit expression of force calculation}
\label{alg_calcforce_explicit}
\begin{algorithmic}[1]
\FOR{$k = 1$ \TO $N_\mathrm{pair}$}
\STATE $(i,j)$ $\leftarrow$ PairList[$k$]
\STATE $\textbf{r} \leftarrow  \textbf{q}[j] - \textbf{q}[i]$
\STATE $r^2 \leftarrow  |\textbf{r}|^2$
\STATE $r^6 \leftarrow r^2 \times r^2 \times r^2$
\STATE $r^{14} \leftarrow r^6 \times r^6 \times r^2$
\STATE $f dt \leftarrow  \left[(24 r^6 - 48)/ r^{14} + 8 c_2 \right] \times dt$
\STATE $\textbf{p}[i] \leftarrow \textbf{p}[i] + fdt \times \textbf{r}$
\STATE $\textbf{p}[j] \leftarrow \textbf{p}[j] - fdt  \times \textbf{r}$
\ENDFOR
\end{algorithmic}
\end{algorithm}
As shown in Algorithm~\ref{alg_calcforce_explicit}, 
the force calculation of the Lennard-Jones potential involves at least one division operation for each pair.
Divisions are sometimes expensive compared with multiplications.
Therefore, computations can be made faster by reducing the number of divisions. 
Suppose there are two independent divisions as follows;
\begin{algorithmic}
\STATE $A_1 \leftarrow 1/B_1$,
\STATE $A_2 \leftarrow 1/B_2$.
\end{algorithmic}
We can reduce the number of divisions by transforming them into
\begin{algorithmic}
\STATE $C \leftarrow 1/(B_1 \times B_2)$,
\STATE $A_1 \leftarrow C \times B_2$,
\STATE $A_2 \leftarrow C \times B_1$.
\end{algorithmic}
Here, the number of divisions decreases from two to one, while three multiplications appear.
This optimization can be effective for architecture in which the penalty for divisions is large.
We apply this optimization technique to the force calculation of the Lennard-Jones system
by utilizing loop unrolling, that is, we first obtain two independent divisions by unrolling and then we apply
this method of reducing the number of divisions to the unwound loop.
Although one can apply this optimization technique to Algorithm~\ref{alg_calcforce_explicit},
it is more efficient to use it together with the sorted array described in Sec.~\ref{sec_sorted_pair list}.
The algorithm for the force calculation with a reduced number of divisions is shown in Algorithm~\ref{alg_division_reduction}.
The term $\left\lfloor x \right\rfloor$ denotes the largest integer less than or equal to $x$ and
subscriptions $a$ and $b$ correspond to the loop unrolled twice.
This optimization achieves 10\% speedup in the condition described in Sec.~\ref{sec_benchmarkresults}.
 
\begin{algorithm}
\caption{Calculating the force using the sorted array and reduction of divisions}
\label{alg_division_reduction}
\begin{algorithmic}[1]
\FOR{$i$ = 1 \TO $N-1$}  
\STATE $\textbf{q}_\mathrm{key} \leftarrow \textbf{q}[i]$
\STATE $\textbf{f}_\mathrm{key} \leftarrow 0$
\STATE $n \leftarrow $ KeyPointer[$i+1$] - KeyPointer[$i$]
\FOR{$k = 1$ \TO $\lfloor n/2 \rfloor$} 
\STATE $k_a \leftarrow $ KeyPointer[$i$] + $2(k-1)$
\STATE $k_b \leftarrow $ KeyPointer[$i$] + $2(k-1)+1$
\STATE $j_a \leftarrow $ SortedList[$k_a$]
\STATE $j_b \leftarrow $ SortedList[$k_b$]
\STATE $\textbf{r}_a \leftarrow \textbf{q}[j_a] - \textbf{q}_\mathrm{key} $ 
\STATE $\textbf{r}_b \leftarrow \textbf{q}[j_b] - \textbf{q}_\mathrm{key} $ 
\STATE $r^2_a \leftarrow |\textbf{r}_a|^2 $ 
\STATE $r^2_b \leftarrow |\textbf{r}_b|^2 $ 
\STATE $r^6_a \leftarrow r^2_a \times r^2_a \times r^2_a$
\STATE $r^6_b \leftarrow r^2_b \times r^2_b \times r^2_b$
\STATE $r^{14}_a \leftarrow r^6_a \times r^6_a \times r^2_a$
\STATE $r^{14}_b \leftarrow r^6_b \times r^6_b \times r^2_b$
\STATE $D \leftarrow 1/ (r^{14}_a \times r^{14}_b)$
\STATE $fdt_a  \leftarrow  \left[ (24 r^6_a - 48) \times D \times r^{14}_b + 8 c_2 \right] \times dt$
\STATE $fdt_b  \leftarrow  \left[ (24 r^6_b - 48) \times D \times r^{14}_a + 8 c_2 \right] \times dt$
\STATE $\textbf{f}_\mathrm{key} \leftarrow \textbf{f}_\mathrm{key} + fdt_a \times \textbf{r}_a + fdt_b \times  \textbf{r}_b$
\STATE $\textbf{p}[j_a] \leftarrow \textbf{p}[j_a] - fdt_a \times \textbf{r}_a$
\STATE $\textbf{p}[j_b] \leftarrow \textbf{p}[j_b] - fdt_b \times \textbf{r}_b$
\ENDFOR
\IF{ $n$ is odd} 
\STATE calculate force for the last partner particle.
\ENDIF
\STATE $\textbf{p}[i] \leftarrow \textbf{p}[i] + \textbf{f}_\mathrm{key}$
\ENDFOR
\end{algorithmic}
\end{algorithm}

\subsection{Cache Efficiency}

\begin{figure}[tb]
\begin{center}
\includegraphics[width=0.7\linewidth]{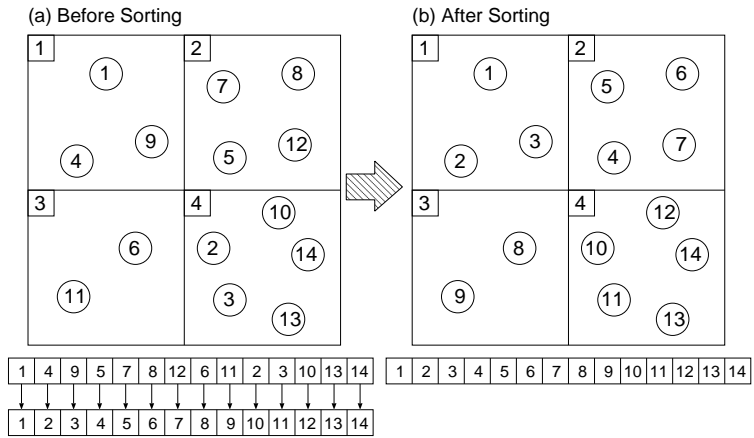}
\end{center}
\caption{ Sorting the indices of the particles to increase cache efficiency. There are four cells and 14 particles in the system.
(a) Status of the system before sorting. The array of grid information corresponding to this configuration
is shown below (see Fig.~\ref{fig_gridlist}). (b) Indices of the particles are sorted using the grid information.
}
\label{fig_sort}
\end{figure}

As particles move, the indices of the partner particles of each key-particle become random.
Then the data of the interacting particles, which are spatially close, may be widely separated in memory. 
This severely decreases the computational efficiency since the data of particles that are not stored in the cache are frequently required.
To improve the situation, reconstruction of the order of the particle indices in arrays is proposed~\cite{Meloni2007}.
This involves sorting so that the indices of interacting particles are as close together as possible.
This can be implemented by using the grid information described in Sec.~\ref{sec_grid_search}.
After an array of grid information is constructed, the particles are arranged in the order in the array.
The sorting algorithm is shown in Fig.~\ref{fig_sort}.
After sorting, the indices of the particles in the same cell become sequential, which improves the cache efficiency.
Note that the pair list is made invalid by this procedure; therefore, 
the pair list construction and the sorting should be performed at the same time.

\begin{figure}[tb]
\begin{center}
\includegraphics[width=0.7\linewidth]{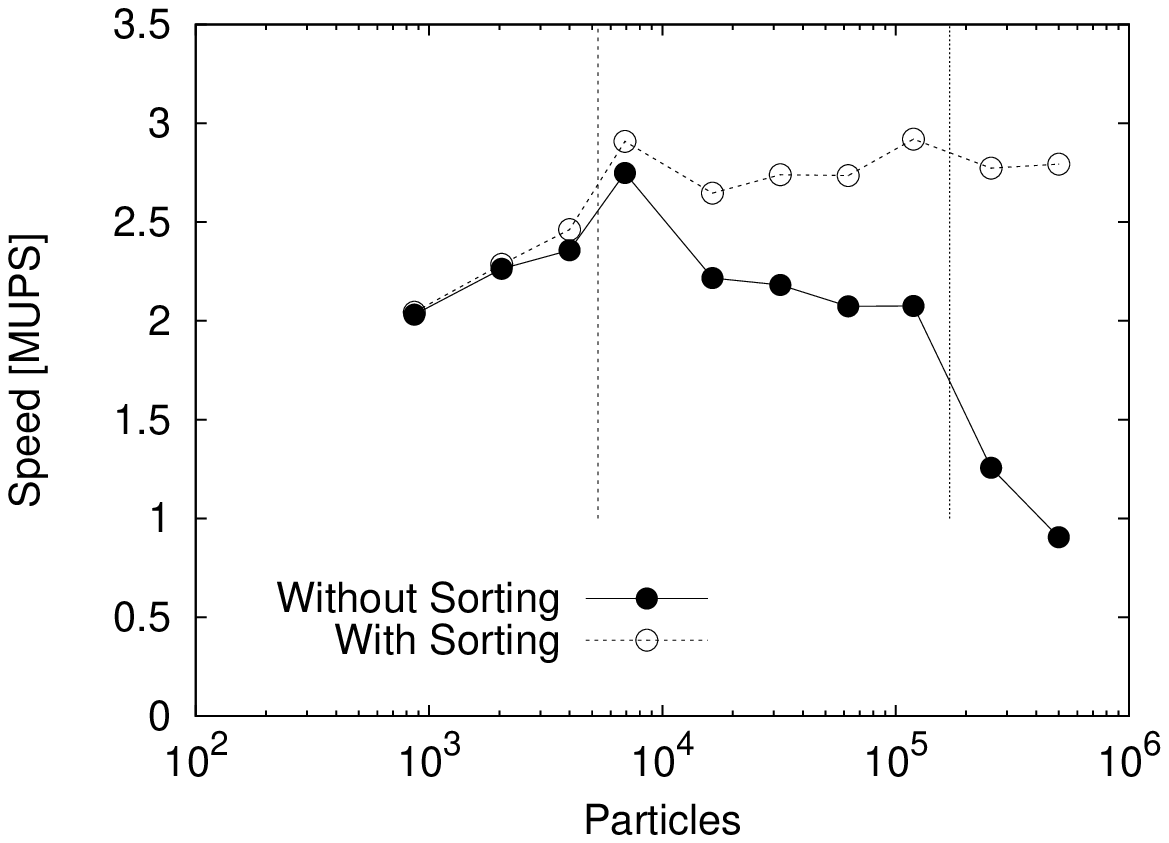}
\end{center}
\caption{ Computational speed with and without sorting. The speed is greatly improved by sorting for a large number of particles $N$.
Dashed vertical lines denote $N = 5.3 \times 10^3$ and $N \sim 1.7 \times 10^5$.
The behavior changes at the lines reflecting the cache hierarchy.
}
\label{fig_sortresult}
\end{figure}

A comparison between the computational speed with and without sorting is shown in Fig.~\ref{fig_sortresult}.
The simulations were performed on a Xeon 2.93 GHz machine with 256 KB for an L2 cache and 8 MB for an L3 cache.
We performed the sorting every ten constructions of a pair list, which is typically every few hundred time steps,
since the computational cost of sorting is not expensive but not negligible.
To mimic a configuration after a very long time evolution without sorting,
the indices of the particles were completely shuffled at the beginning of the simulation, 
that is, the labels of the particles were exchanged randomly while keeping their positions and momenta.
The computational speed is measured in the unit of MUPS (millions update per second),
which is unity when one million particles are updated in one second.
As the number of particles increased, the computational efficiency was greatly improved by sorting.
The stepwise behavior of the data without sorting reflected the cache hierarchy.
Since the system is three dimensional, amount of memory required to store information of one particle is 48 bytes.
Therefore, L2 cache (256) KB corresponds to 256 KB/48 bytes $\sim 5.3 \times 10^3$ and
L3 cache (8MB) corresponds 8 MB/48 bytes $1.7 \times 10^5$.
When the number of particles was smaller than $5.3 \times 10^3$,
the difference in speed between the data with and without sorting was moderate.
While the number of particles was larger than $5.3 \times 10^3$, 
we found that the efficiency was improved by 30 to 40\% by sorting.
When the number of particles was larger than $1.7 \times 10^5$,
the computational speed without sorting became much lower than that with sorting.
We found that the effect of the sorting in efficiency was about a factor of 3 in the case of $10^6$ particles.

\subsection{Software Pipelining}

\begin{figure}[tb]
\begin{center}
\includegraphics[width=0.99\linewidth]{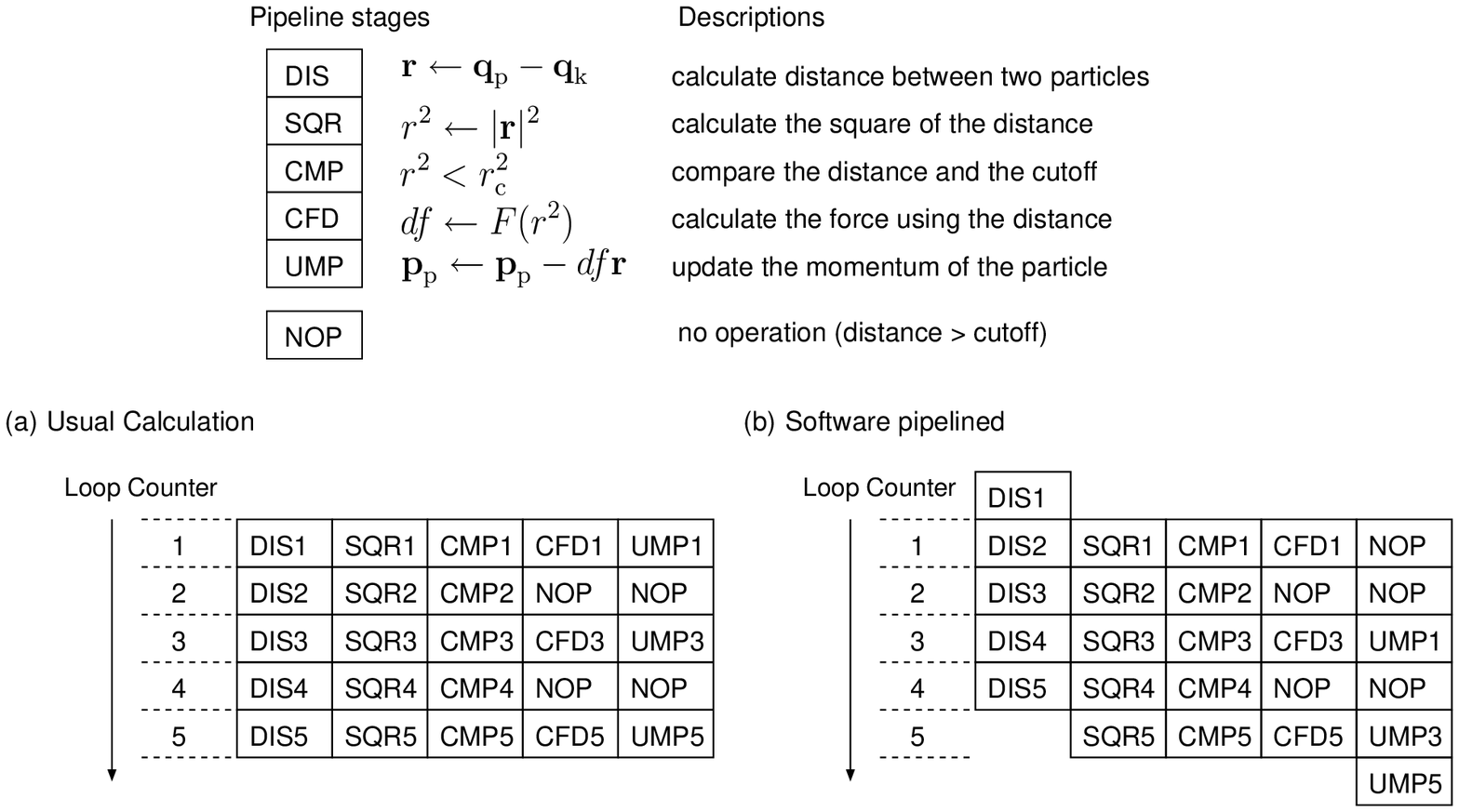}
\end{center}
\caption{ Software pipelining. 
Calculation of the force consists of five parts: DIS, SQR, CMP, CFD, and UMP.
(a) Implementation without software pipelining. All calculation are performed sequentially.
(b) Implementation with software pipelining.
The order of the instructions are arranged in order to improve the throughput of the instructions.
}
\label{fig_pipeline}
\end{figure}

\begin{algorithm}
\caption{Calculating the force with software pipelining}
\label{alg_pipeline}
\begin{algorithmic}[1]
\FOR{$i= 1$ \TO $N-1$}  
\STATE $\textbf{q}_\mathrm{key} \leftarrow \textbf{q}[i]$
\STATE $\textbf{p}_\mathrm{key} \leftarrow \textbf{0}$
\STATE $fdt \leftarrow 0$
\STATE $k \leftarrow $ KeyPointer[$i$]
\STATE $j_a \leftarrow$ SortedList[$k$]
\STATE $\textbf{r}_a \leftarrow \textbf{q}[j_a] - \textbf{q}_\mathrm{key}$
\STATE $\textbf{r}_b \leftarrow \textbf{0}$
\STATE $j_b \leftarrow 0$
\FOR{$k=$ KeyPointer[$i$] \TO KeyPointer[$i+1$]  -1 } 
\STATE $\textbf{r} \leftarrow \textbf{r}_a$
\STATE $r^2 \leftarrow |\textbf{r}|^2$
\STATE $j \leftarrow j_a$
\STATE $j_a \leftarrow \mathrm{SortedList}[$k+1$]$
\STATE $\textbf{r}_a \leftarrow \textbf{q}[j_a] - \textbf{q}_\mathrm{key}$
\IF {$r^2 < r_\mathrm{c}^2$}
\STATE $\textbf{p}_\mathrm{key} \leftarrow \textbf{p}_\mathrm{key} + fdt \times \textbf{r}_b$
\STATE $\textbf{p}[j_b] \leftarrow \textbf{p}[j_b] - fdt \times \textbf{r}_b$
\STATE $fdt \leftarrow [ (24 r^6-48)/ r^{14} + 8c_2] \times dt$ 
\STATE $j_b \leftarrow j$
\STATE $\textbf{r}_b \leftarrow \textbf{r}$
\ENDIF
\ENDFOR
\STATE $\textbf{p}[j_b] \leftarrow \textbf{p}[j_b] - fdt \times \textbf{r}_b$
\STATE $\textbf{p}[i] \leftarrow \textbf{p}[i] + \textbf{p}_\mathrm{key}  + fdt \times \textbf{r}_b$
\ENDFOR
\end{algorithmic}
\end{algorithm}
The elimination of the conditional branch can be inefficient in some types of CPU architectures such as Intel or AMD,
since the cost of the conditional branch is not expensive. In such CPU architectures, software pipelining is 
effective in improving the computational speed.
Recent CPUs can handle four or more floating-point operations simultaneously if they are independent.
The force calculation of one pair, however, consists of a sequence of instructions that are dependent on each other.
Therefore, CPU sometimes has to wait for the previous instruction to be completed, which decreases the computational speed.
Software pipelining is a technique to reduce such dependencies between instructions
in order to increase the efficiency of the calculation.
The purpose of the software pipening is increasing a number of instructions which can be
executed simultaneously, and consequently, increasing the instruction throughputs
in a hardware pipepine.
While the software-pipelining technique is often used in combination with loop unrolling,
the simple loop unrolling can cause a shortage of registers
leading to access to the memory which can be expensive.
The main idea of software pipelining is to calculate independent parts of the force calculation simultaneously by shifting the order of the instructions.
The calculation of force can be broken into the following five stages;\\
\textbf{DIS} calculate distance between two particles. \\
\textbf{SQR} calculate the square of the distance. \\
\textbf{CMP} compare the distance and the cutoff length. \\
\textbf{CFD} calculate the force using the distance.  \\
\textbf{UMP} update the momentum of the particle. \\
The schematic image of the software pipelining is shown in Fig.~\ref{fig_pipeline}.
The number after a stage denotes the index of the pair (the value of the loop counter), \textit{e. g.},
DIS1 denotes the calculation of the distance of the first particle pair.
Figure~\ref{fig_pipeline}~(a) denotes the calculation without software pipelining.
Suppose the distances between third and fifth pairs are longer than the cutoff length.
Then the calculation of the force and the update the momentum are unnecessary.
We refer this to \textbf{NOP} (no operation).
All calculations are performed sequentially. For instance, SQR1 is performed after
DIS1 is completed, CMP1 is performed after SQR1 is completed, and so on.
Figure~\ref{fig_pipeline}~(a) denotes the calculation with software pipelining.
The order of the calculations are shifted so that the number of independent calculations
are increased. For instance, DIS4, SQR3, and UMP1 can be performed simultaneously
since there are no dependencies between them (see the column where the value of the loop counter is three).
Increase of independent instructions increases the efficiency of the hardware pipeline,
and consequently, impoves the computational speed
The pseudo code corresponding to Fig.~\ref{fig_pipeline} is shown in Algorithm~\ref{alg_pipeline}.
As far as we know in our experience, the software pipelining without loop unrolling described
here is the best choice for recent architecture such as Intel Xeon processors.

\section{Parallelization}
\label{sec_parallelization}

\subsection{Parallelization Scheme}

\begin{figure}[tb]
\begin{center}
\includegraphics[width=0.9\linewidth]{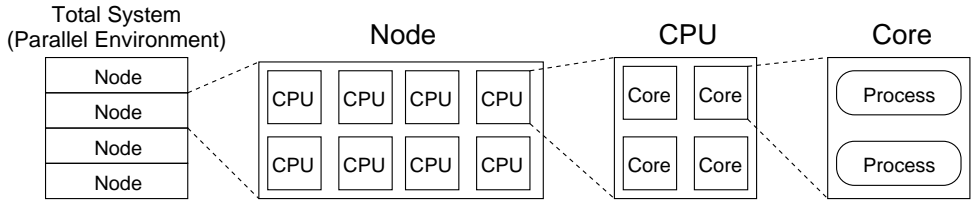}
\end{center}
\caption{ Parallel environment. A total system of a supercomputer consists of 
several nodes. A node consists of several CPUs. A CPU consists of several cores.
One or two processes are executed on a core.
}
\label{fig_p_environment}
\end{figure}

Since the size of the system or computational steps are limited on the single process job,
parallel computation is required in order to simulate large systems or long time steps.
Parallel computation is executed on a parallel environment such as a supercomputer.
A total system of a supercomputer has hierarchic structure, \textit{i.e.},
a total system consists of nodes, a node consists of CPUs, a CPU consists of cores (see Fig.~\ref{fig_p_environment}).
One or two processes is usually executed on a single core, and a set of processes
work concertedly on the same task.

There has been considerable effort devoted to the parallelization of MD simulations.
The parallel algorithms used for MD can be classified into three strategies, domain decomposition, particle decomposition, and force decomposition~\cite{Plimpton1995}.
In domain decomposition (also called spatial decomposition)
the simulation box is subdivided into small domains, and each domain is assigned to each process.
In particle decomposition, the  computational workload is divided and distributed 
on the basis of the particles. Different processes have different particles.
In force decomposition the computational workload is divided and distributed 
on the basis of the force calculation. Consider a matrix $F_{ij}$ that denotes the force between
$i$- and $j$-particles. Force decomposition is based on the block decomposition of this force matrix.
Each block of the force matrix is assigned to each process.
Generally, domain decomposition is suitable for systems with short-range interactions
and simulations with a large number of particles~\cite{Hayashi, Beazley}, and force decomposition is suitable for systems
with long-range interactions such as electrostatic force and simulations with a large number of steps.
In the present paper, we adopt a simple domain decomposition method for parallelization, \textit{i.e.},
we divide a system into small cubes of identical size.
We use a Message Passing Interface (MPI) for all communication.
To perform the computation, three types of communication are necessary:
(i) exchanging particles that stray from the originally associated process,
(ii) exchanging information of particles near boundaries to consider the cross-border interactions, and 
(iii) synchronization of the validity check of pair lists.

\subsection{Exchanging Particles}

Although all particles are initially in the domains to which they are placed, they tend to stray from their original domains
as a simulation progresses. Therefore, each process should pass the migrated particles
to an appropriate process at an appropriate time.
There is no need for exchanging particles each step
since it will not cause problems while migration distances are short.
Therefore, migrated particles are exchanged simultaneously with a reconstruction of pair list.

\subsection{Cross-Border Interactions}
\label{sec_cross_border}

\begin{figure}[tb]
\begin{center}
\includegraphics[width=0.95\linewidth]{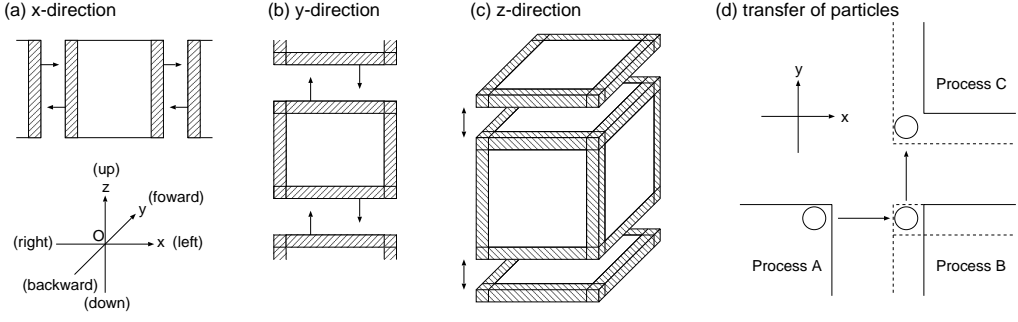}
\end{center}
\caption{
Sending and receiving of positions of particles that are close to boundaries.
(a) Sending along the $x$-direction. First, each process sends/receives the positions of particles near the right boundary to/from 
the right neighbor. Similarly, the positions of particles near the left boundary is transmitted.
(b) Sending along the $y$-direction. Positions of particles near the boundaries is sent to neighbors.
(c) Sending along the $z$-direction is performed similarly.
(d) Transfering particles. Suppose there are three neighboring domains associated with Process A, B, and C.
First Process A sends a particle to Process B along $x$-axis. Process B sends the particle recieved from Process A
to Process C. Then Process C recieves particles from Process A via Process B.
Thus, each process obtains the information of particles from its diagonal neighbors without direct communication.
}                                                                                                                                                                                                                                                                                                                                                                              
\label{fig_sendblock}
\end{figure}

To calculate forces between particle pairs which are assigned to different processes,
communication between neighboring processes is necessary.
While a naive implementation involves communication in 26 directions since each process has 26 neighbors,
the number of communications can be reduced to six by forwarding information of particles sent by other processes.
Suppose a process is assigned the cuboidal domain $s_x \leq x < e_x, s_y \leq y < e_y$, and $s_z \leq z < e_z$,
and the search length is denoted by $r_\mathrm{s}$, which is defined in Sec.~\ref{sec_dutc}.
The procedures to send and receive positions of particles on the boundaries are as follows.
\begin{algorithmic}[1]
\STATE Left: Send  positions of particles in the region $s_x \leq x < s_x + r_\mathrm{s}$ to the left process, and
receive positions of particles in the region $s_x - r_\mathrm{s} \leq x < s_x$ from the left process.
\STATE Right: Send the positions of particles in the region $e_x - r_\mathrm{s} \leq x < e_x$, and
receive positions of particles in the region $e_x \leq x < e_x + r_\mathrm{s}$ from the right process.
\STATE Backward: Send  positions of particles in the region $s_y \leq y < s_y + r_\mathrm{s}$ to the backward process,
including the particles received from the left and the right process.
Receive positions of particles in the region $s_y - r_\mathrm{s} \leq y < s_y$ from the backward process.
\STATE Forward: Send the positions of particles in the region $e_y - r_\mathrm{s} \leq y < e_y$ to the process in front,
including the particles received from left and the right processes.
Receive positions of particles in the region $e_y \leq y < e_y + r_\mathrm{s}$ from the process in front.
\STATE Down: Send  the positions of particles in the region $s_z \leq z < s_z + r_\mathrm{s}$ to the lower process,
including the particles received from other processes.
Receive positions of particles in the region $s_z - r_\mathrm{s} \leq z < s_z$ from the lower process.
\STATE Up: Send positions of particles in the region $e_z - r_\mathrm{s} \leq z < e_z$ from the upper process,
including the particles received from other processes.
Receive positions of particles in the region $e_z  \leq z < e_z + r_\mathrm{s}$ from the upper process.
\end{algorithmic}
After the above six communications, exchanging positions of particles close to boundaries
are completed including between diagonal processes.
The details are illustrated in Fig.~\ref{fig_sendblock}.
Note that the simple implementation of the communication may cause deadlock
which is a situation that two or more processes will wait responce forever.
Suppose there are two processes, A and B.
If A tries to send a message to B, and B also tries to send to A with MPI blocking communications,
then the communication will not be completed.
Similarly, suppose there are three processes, A, B, and C.
If the communication path are A $\rightarrow$ B, B $\rightarrow$ C, and 
C $\rightarrow$ A, then the system is getting into a deadlock.
In the domain decomposition parallelism with periodic boundary condition,
it is possible that a deadlock occurs.
There are several techniques to avoid this problem, such as 
reversing the order of sending and receiving and using nonblocking operations.
A simple way to avoid such a deadlock problem is to use the MPI\_Sendrecv function.
This function does not cause deadlock for closed-path communications.
The pseudocode using MPI\_Sendrecv is shown in Algorithm~\ref{alg_sendrecv}.
$SendDir$ and $RecvDir$ denote the direction that the process should send to or recieve from.
There are six directions to complete the communication.
The number of particles to be sent in each direction is fixed until the expiration of a pair list.
Suppose $N_\mathrm{s}$ is a number of particles to send.
Then the number of data sent is $3 N_\mathrm{s}$, since only coordinates are required to calculate forces.
The message tag denoted by $Tag$, which is an arbitrary non-negative integer but it should be
the same for sender and receiver.

\begin{algorithm}
\caption{Communications for cross border interactions}
\label{alg_sendrecv}
\begin{algorithmic}[1]
\FOR{$SendDir$ in [left, right, backward, forward, down, up]}  
\STATE $RecvDir \leftarrow$ the opposite direction of $SendDir$
\STATE $SendBuf \leftarrow$ coordinates of particles to send
\STATE $N_\mathrm{s} \leftarrow$ number of particles to send
\STATE $N_\mathrm{r} \leftarrow$ number of particles to receive
\STATE Prepare $RecvBuf$
\STATE $DestRank \leftarrow $ rank of neighbor on the $SendDir$.
\STATE $SrcRank \leftarrow $ rank of neighbor on the $RecvDir$.
\STATE Call MPI\_Sendrecv(%
$SendBuf$, $3 N_\mathrm{s}$, MPI\_DOUBLE, $DestRank$, $Tag$, 
$RecvBuf$, $3 N_\mathrm{r}$, MPI\_DOUBLE, $SrcRank$, $Tag$, MPI\_COMM\_WORLD)
\STATE Update coordinates of particles using $RecvBuf$
\ENDFOR
\end{algorithmic}
\end{algorithm}

\subsection{Synchronization of pair list Constructions}

\begin{figure}[tb]
\begin{center}
\includegraphics[width=0.8\linewidth]{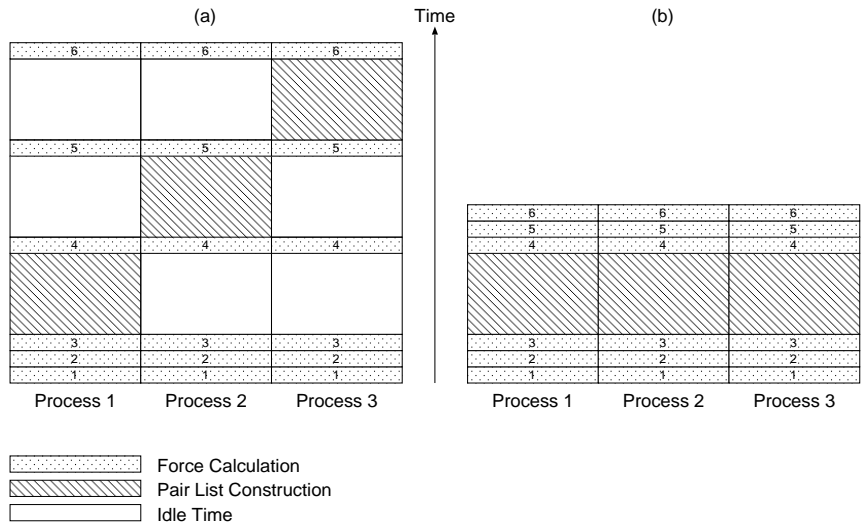}
\end{center}
\caption{
Synchronization of pair lists.
The numbers in the boxes of force calculation denote the time steps.
The size of the boxes of force calculation and the pair list construction reflect
the computational time required for them, \textit{i.e.}, the size of the box
is proportional to the computational time.
Suppose the lifetime of the pair list of Process 1 is 3 steps, that is, 
the pair list should be reconstructed after three force calculations.
Similarlly, the lifetime of the pairlists are 4 and 5 for Process 2 and 3,  respectively.
(a) Without synchronization, if the processes reconstruct pair lists independently, then other processes must wait 
for the reconstruction resulting in a huge amount of idle time.
(b) With synchronization, if the processes synchronize their own pair lists with others, then
the idle time vanishes while the average lifetime of pair lists also decreases.
}
\label{fig_pair list}
\end{figure}

Each process maintains its own pair list and has to reconstruct it when it expires.
If each process checks the validity of the pair list and reconstructs it independently,
since it causes idle time, and the calculation effiency dicreases.
To solve this problem, we synchronize the validity check of pair lists for all processes,
that is, we reconstruct all the pair lists even when one of the pair lists has expired.
While this reduces the average lifetime of the pair lists, the overall speed of the simulation
is greatly improved since the idle time is eliminated (see Fig.~\ref{fig_pair list}).
This synchronization process can be implemented with MPI\_Allreduce,
which is a global-reduction function of the MPI.
Note that the global synchronization of pair lists can have a serious effect on the parallel efficiency
when a system is highly inhomogeneous such as a system involving phase transitions
or a heat-conducting system. 
In such cases, further advanced treatment that manages pair lists locally at each location of grid would be required.

\subsection{Results of Benchmark Simulations} \label{sec_benchmarkresults}

\begin{figure}[tb]
\begin{center}
\includegraphics[width=0.8\linewidth]{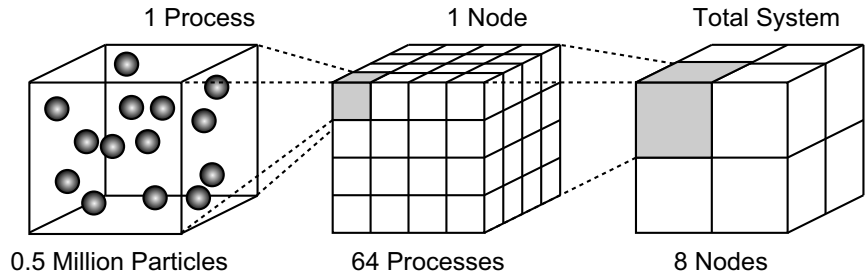}
\end{center}
\caption{
Domain decomposition for parallelization at NIFS.
Each process is assigned a cube with a length of $100 $ that contains 500,000 particles.
Each node is assigned to a large cube made up of $4\times 4\times 4 $ small cubes.
The case of eight nodes is shown as an example.
}
\label{fig_blocks}
\end{figure}

To measure the efficiency of the developed code, we perform benchmark simulations on two different parallel computers located at the different facilities, which are NIFS and ISSP.
The details of the computers are listed in Table~\ref{tbl_computers}.
In the actual analysis we used the so-called week scaling analysis,
which is the dependence of the execution time in terms of node number
under the condition of the number of particle per process keeping constant.
The conditions for the benchmark simulations are as follows.
 Note that, all quantities are normarized by well depth $\varepsilon$ and atomic diameter $\sigma$
of the Lennards-Jones potential defined in Sec.~\ref{sec_potential}.
\begin{itemize}
\item System size: $100 \times 100 \times 100$ for each process.
\item Number of particles: 500,000 for each process (number density is 0.5).
\item Initial condition: Face-centered-cubic lattice.
\item Boundary condition: Periodic for all axes.
\item Integration scheme: Second-order symplectic integration.
\item Time step: 0.001.
\item Cutoff length: 2.5.
\item Search length: 2.8. 
\item Cutoff scheme: Add constant and quadratic terms to potential as shown in Eq.~(\ref{eq_lj_cutoff}).
\item Initial velocity: The absolute value of the velocities of all particles are set to be 0.9 and
 the directions of the velocities are given randomly.
\item After 150 steps, we measure the calculation (elapsed) time for the next 1000 steps.
\end{itemize}
When a system with $N$ particles is simulated for $k$ steps in $t$ seconds, then
the number of MUPS is given by $ \times Nk/10^{6}t$.
At NIFS, we placed $4\times 4 \times 4$ cubes at each node, and assigned one cube to each process (see Fig.~\ref{fig_blocks}),
since each node has 32 cores, and we placed 64 processes on each node utilizing simultaneous multithreading (SMT).
At ISSP, we placed $2\times 2 \times 2$ cubes at each node, and assigned one cube to each process, since one node consists of two CPUs (8 cores) at ISSP.

\begin{table}[tb]
\begin{tabular}{llllll}
\hline
Facility &  Name                              &  CPU                                           &  Cores  & Processes &  Memory  \\\hline
NIFS         & HITACHI SR16000/J2   &  IBM POWER6 (4.7 GHz)            & 32                   & 64 (*) &  128GB                \\
ISSP          & SGI Altix ICE 8400EX    &  Intel Xeon X5570 (2.93 GHz) & 8                     &  4 or 8 (**) &   24GB                \\\hline
\end{tabular}
\caption{ Summary for
HITACHI SR16000/J2 at the National Institute for Fusion Science (NIFS) and
SGI Altix ICE 8400EX at the Institute for Solid State Physics (ISSP).
Location, name, CPU, number of cores per node, number of processes per node, and memory per node are shown.
(*) Using simultaneous multithreading (SMT), 64 processes are executed on each node.
(**) The single-node run is performed with four processes on one CPU and two-or-more-node runs are performed 
with 8 processes per node on two CPUs because of the queue design at ISSP.
}
\label{tbl_computers}
\end{table}

\begin{figure}[tb]
\begin{center}
\includegraphics[width=0.9\linewidth]{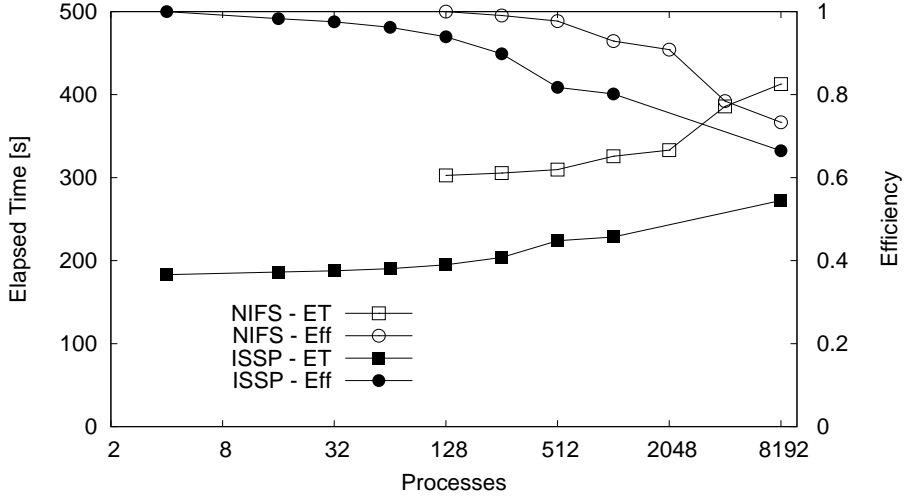}
\end{center}
\caption{ Results of benchmark simulations.
The results at NIFS are denoted by open symbols and ISSP by filled symbols.
ET and Eff denotes elapsed time [s] and efficiency, respectively.
The increase in elapsed time is due to the parallel overhead.
}
\label{fig_results}
\end{figure}

The results obtained at NIFS and ISSP are summarized in Table~\ref{tbl_results_NIFS} and Table~\ref{tbl_results_ISSP}, respectively.
Both results are shown in Fig.~\ref{fig_results}.
While the elapsed time of the 2-node calculation at NIFS is 302.6 s, 
that of the 128-node calculation  is 412.7 s.
Similarly, while the elapsed time of the single node calculation at ISSP is 183.13 s,
that of the 1024-node calculation is 272.478 s.
Since we perform weak scaling, the elapsed time should be independent of the number of processes
provided the computations are perfectly parallelized.
Therefore, the increase in elapsed time of the calculation with larger nodes is regarded as being due to the parallel overhead.

\begin{table}[htb]
\begin{center}
\begin{tabular}{rrrrrc}
\hline
Nodes & Processes & Particles & Elapsed Time [s] & Speed [MUPS] & Efficiency\\\hline
2 & 128 & 64,000,000 & 302.623 & 211.484 & 1.00 \\
4 & 256 & 128,000,000 & 305.469 & 419.027 & 0.99 \\
8 & 512 & 256,000,000 & 309.688 & 826.638 & 0.98 \\
16 & 1024 & 512,000,000 & 325.744 & 1571.79 & 0.93 \\
32 & 2048 & 1,024,000,000 & 333.140 & 3073.78 & 0.91 \\
64 & 4096& 2,048,000,000 & 385.765 & 5308.93 & 0.78 \\
128 & 8192 & 4,096,000,000 & 412.717 & 9924.48 & 0.73 \\
\hline
\end{tabular}
\end{center}
\caption{
Results of benchmark simulations at NIFS. Numbers of nodes, numbers of processes, numbers of particles, elapsed times [s], and 
speeds [MUPS] are listed.
Efficiency is estimated based on the elapsed time obtaind by the case of the 2-node calculation, 
since a single-node calculation is not permitted at NIFS.
}
\label{tbl_results_NIFS}
\end{table}

\begin{table}[htb]
\begin{center}
\begin{tabular}{rrrrrc}
\hline
Nodes & Processes & Particles & Elapsed Time [s] & Speed [MUPS] & Efficiency\\\hline
1 & 2000000 & 183.13 & 10.9212 & 1.00 \\
2 & 8000000 & 186.296 & 42.9424 & 0.98 \\
4 & 16000000 & 187.751 & 85.2194 & 0.98 \\
8 & 32000000 & 190.325 & 168.133 & 0.96 \\
16 & 64000000 & 194.932 & 328.32 & 0.94 \\
32 & 128000000 & 203.838 & 627.95 & 0.90 \\
64 & 256000000 & 224.028 & 1142.71 & 0.82 \\
128 & 512000000 & 228.485 & 2240.84 & 0.80 \\
1024 & 4096000000 & 272.478 & 15032.4 & 0.66 \\
\hline
\end{tabular}
\end{center}
\caption{
Same as Table~\ref{tbl_results_NIFS} for ISSP.
Efficiency is determined relative to the elapsed time for the single-node calculation.
}
\label{tbl_results_ISSP}
\end{table}

\section{Parallel Overhead}
\label{sec_parallel_overhead}

The benchmark results described in the previous section show that
the  parallelization efficiency decreases as a number of processes increases,
while the ammount of communication in one step seems to be negligible compared with the computations.
In the present section, we investigate the cause of the parallel overhead of our codes
in order to identify the factor of the decrease in parallelization efficiency.
All simulations presented here are performed under the same condition
as that used for the benchmarks in Sec.~\ref{sec_benchmarkresults}.

\subsection{Granularity}

In parallel computing, the ratio of computation to communication is important.
This ratio is called the \textit{granularity}.
Fine-grain parallelism refers to the situation that data are transferred frequently
and computational tasks are relatively light compared with communication.
Coarse-grained parallelism, on the other hand, refers to the situation that computational tasks are relatively heavy
compared with communication. In general,  a program having coarse granularity exhibits
better performance for a large number of processes.
Here, we estimate the granularity of our code.
Although the interaction length is set to $2.5$, each process has to send and receive information of particles
whose distances from boundaries are less than $2.8$, since we adopt the DUTC method and
the search length is set to $2.8$.
Each process is assigned to a cube of length $100$ and the number density is set to $0.5$.
The communications are performed along the $x$-, $y$-, and $z$-directions, and 
two communications are involved for each direction.
The amount of transmitting data increases in order of the $x$-, $y$-, and $z$- directions, since
the received data of particles are forwarded.
The position of each particle is expressed by three double-precision floating-point numbers, requiring $3\times 8 =24$ bytes.
The amount of communication along $y$-axis is larger than that along $x$-axis, since the
information of particles recieved along $x$-axis are transfered to $y$-axis (see Fig.~\ref{fig_sendblock}).
The number of particles transmitted along the $x$-axis is $100\times 100 \times 2.8 \times 0.5 =14,000$.
Therefore, 336 KB are transmitted along the $x$-axis.
Similarly, $(100 +2.8\times 2) \times 100 \times 2.8 \times 0.5 \times 24$ = 355 KB of data are transmitted along the $y$-axis
and 375 KB of data are transmitted along the $z$-axis.
Since inner-node communication is performed on shared memory
and should be sufficiently faster than internode communication,
we only consider  the internode communication.
Each node at NIFS has $4\times 4 \times 4 =64$ cubes, and therefore, 16 processes involve internode communication.
The largest amount of communication is along the $z$-axis, which is about 375 KB $\times 16 = $ 6 MB.
The bandwidth between each node at NIFS is about 5 GB/s.
The total amount of internode communication is at most 36 MB, and therefore,
the time spent on communication will be about 7.2 ms for each step.
The computational time without communication is about 0.3 s, which is 500 times longer than the estimated communication time.
This implies that this code has coarse-grained parallelism and should be less affected by increasing the number of processes.
The benchmark results, however, show a significant decrease in parallel efficiency for large-scale runs.

We have also investigated the communication set up time at Kyushu University.
The supercomputer at Kyushu University is HITACHI SR16000/J2 which is same as
that at NIFS. But it is 42 nodes while NIFS has 128 nodes.
We observed the communication set up time for MPI\_Barrier and
MPI\_Allreduce over full of nodes, and found that the set up times are at most 50 us and 130 us, respectively.
Therefore, communication cannot be a main factor of the parallel overhead.

\begin{table}[tbh]
\begin{center}
\begin{tabular}{rrrrc}
\hline
Nodes  &Total [s]   & Force [s] & pair list [s] & Communication [s] \\\hline
2 & 304.609 & 217.403 & 38.26566 & 12.54 \\
128 & 404.335 & 217.733 & 41.85143 & 14.39 \\
\hline
\end{tabular}
\end{center}
\caption{
Breakdowns of the computational time at NIFS. 
Total, Force, pair list, and Communication 
denote the total time of computation [s], the time spent on force calculations [s], the time required to construct pair lists [s],
and communication time [s], respectively.
}
\label{tbl_breakdown_NIFS}
\end{table}

\begin{table}[tbh]
\begin{center}
\begin{tabular}{rrrrc}
\hline
Nodes  &Total [s]   & Force [s] & pair list [s] & Communication [s] \\\hline
1 & 180.945 & 144.92 & 14.388425 & 4.12 \\
256 & 228.614 & 144.638 & 16.299467 & 7.92 \\
\hline
\end{tabular}
\end{center}
\caption{
Same as Table~\ref{tbl_breakdown_NIFS} for ISSP.
}
\label{tbl_breakdown_ISSP}
\end{table}

\begin{figure}[tb]
\begin{center}
\includegraphics[width=0.9\linewidth]{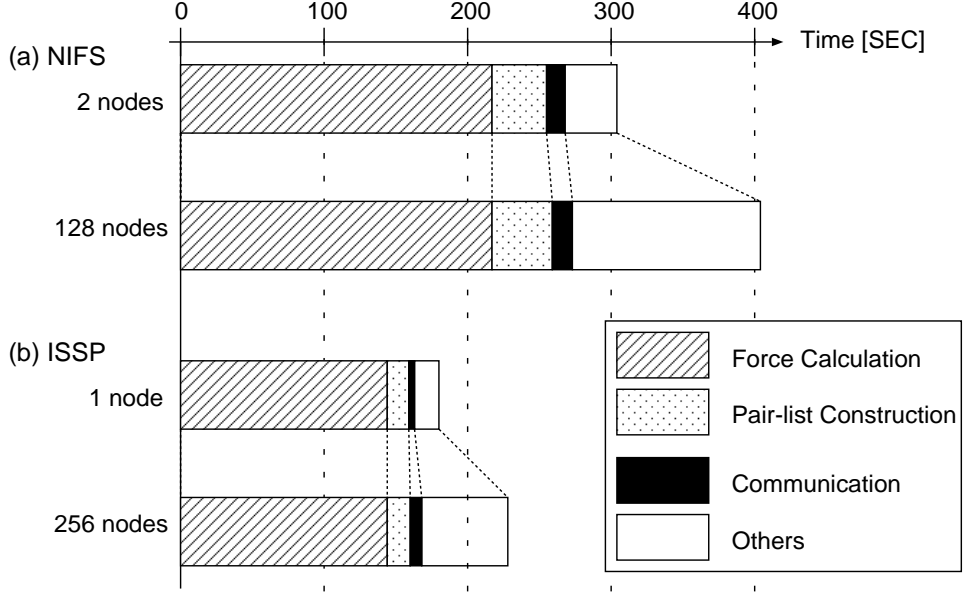}
\end{center}
\caption{Breakdowns of the computational time for (a) NIFS and (b) ISSP.
The smallest and largest runs are shown for both cases.
Although the communication time increases as the number of nodes increases,
the increase in time for communication is at most 8\% of the total increase, which cannot explain the observed parallel overhead.
The idle time due to the synchronization is counted as ``Others".
}
\label{fig_breakdown}
\end{figure}

\subsection{Lifetime of Pair Lists}

Since we perform weak scaling analysis, the size of the system increases as the number of nodes increases.
This can decrease the lifetime of pair lists, since the maximum speed of the particles increases
in a larger system and faster particles reduce the lifetime of pair lists~\cite{MaximumVelocity}.
The construction of a pair list takes much longer time than the force comutation takes, and therefore, 
the simulation run time is strongly affected 
if the number of pair lists being constructed increases.
This increases the simulation run time even if the simulation is perfectly parallelized, and therefore,
it is inappropriate to regard this cost as the parallelization cost.
Therefore, we investigate the system-size dependence on the time required for construction of pair lists.
At NIFS, the pair list was reconstructed 20 times for the 2-node run while
it was reconstructed 21 times for the 128-node run.
The average life time is 51.2 steps for 2-node and 53.8 steps for 128-node caluclations.
Therefore, the computational cost to construct pair lists increases for larger number of nodes, but the difference is small.
The breakdowns of the computational time for the smallest and the largest runs
are listed in Table~\ref{tbl_breakdown_NIFS} and \ref{tbl_breakdown_ISSP} and are also shown in Fig.~\ref{fig_breakdown}.
The additional cost of the pair list construction at NIFS is only 4 [s] which cannot explain the parallel overhead, which is over 100 [s].
Therefore, the lifetime of pair lists is not the main factor causing the reduced parallel performance.

\subsection{Communication Management}

\begin{figure}[tb]
\begin{center}
\includegraphics[width=0.98\linewidth]{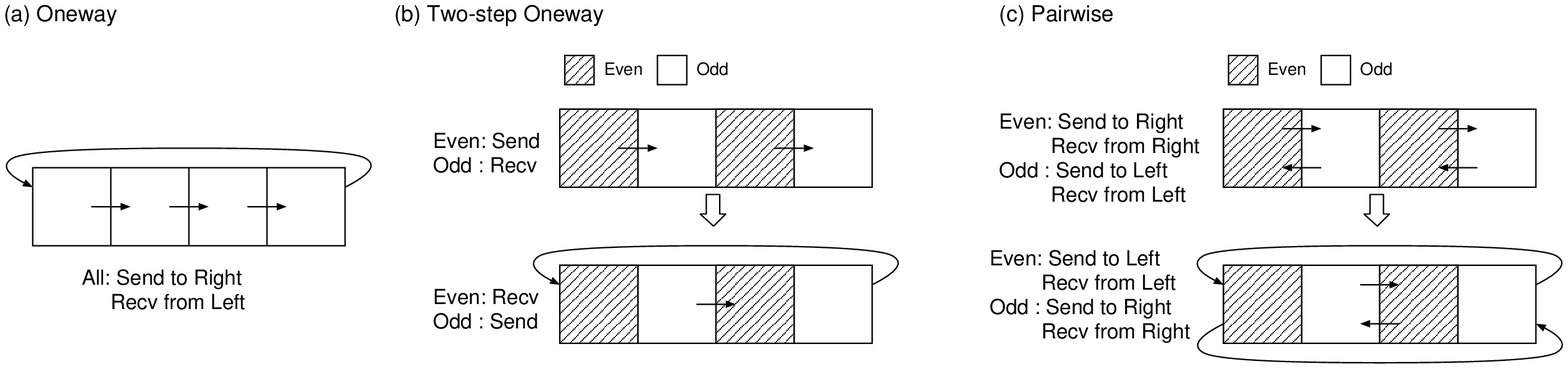}
\end{center}
\caption{ Communication methods.
(a) Oneway: All processes send information in the same direction, and receive the data simultaneously using MPI\_Sendrecv.
(b) Two-step Oneway: Although all processes send information in the same direction, 
the sending and receiving processes are explicitly separated using MPI\_Send and MPI\_Recv. 
(c) Pairwise: Each process is classified as even or odd.
When the even processes exchange the positions of particles with their left neighbors,
the odd processes exchange the positions of particles with their right neighbors.
Then the directions of communication are reversed. Communication is implemented by MPI\_Sendrecv.
}
\label{fig_pairwise}
\end{figure}

Each simulation involves two types of communication at each step.
One is point-to-point communication for sending and receiving data of particles,
the other is collective communication for the synchronization of pair lists.
The point-to-point communication is implemented in a simple way, that is,
all processes send data in one direction and receive data from the opposite direction.
All processes send and receive data simultaneously in the same direction using MPI\_Sendrecv (see Fig.~\ref{fig_pairwise}~(a)).
While the simple implementation of the communication may cause deadlock for the system with THE periodic boundary condition,
the function MPI\_Sendrecv avoid the deadlock implicitly by reordering of communication, 
which may cause some overhead in parallel efficiency for a large number of processes.
Therefore, we examine different ways of communication
which avoid the deadlock explicitly in order to investigate whether the 
communication management by MPI\_Sendrecv is the factor of the parallel overhead.
Here we performed three methods.
The first method, referred to as `Oneway', is the simple implementation described in Sec.~\ref{sec_cross_border}.
The second method, referred to as `Two-step Oneway', involves communications in 2 stages (see Fig.~\ref{fig_pairwise}~(b)).
All processes are classified into two groups, even and odd.
In the first of the two stages, even processes send data and odd processes receive data.
In the second stage, data are transmitted in the reverse direction.
The communications are implemented by the blocking functions MPI\_Send and MPI\_Recv.
The third method, referred to as `Pairwise', arranges the order of communication explicitly. 
All processes are classified into two groups, even and odd, similarly to in the previous method.
When the even processes exchange data with their right neighbors,
the other processes exchange data with their left neighbors, and vice versa (see Fig.~\ref{fig_pairwise}~(c)).
The communications are implemented by MPI\_Sendrecv.
Note that, the behavior of `Pairwise' at ISSP is not monotonic, \textit{i.e.},
the data of 128-node run is much faster than those of 8-, 16-, 32-, and 64-node runs.
This may be due to the network topology at ISSP which is hypercube, however,
the details should be investigated later.

\begin{figure}[tb]
\begin{center}
\includegraphics[width=0.49\linewidth]{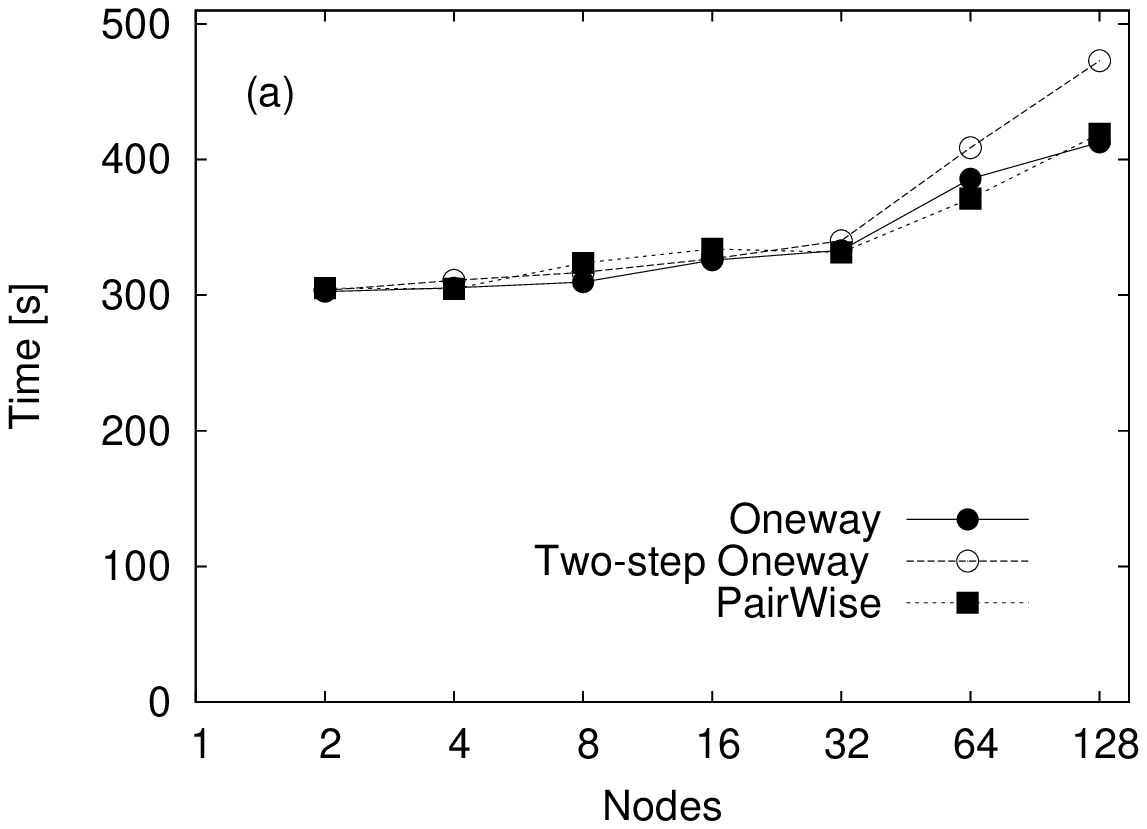}
\includegraphics[width=0.49\linewidth]{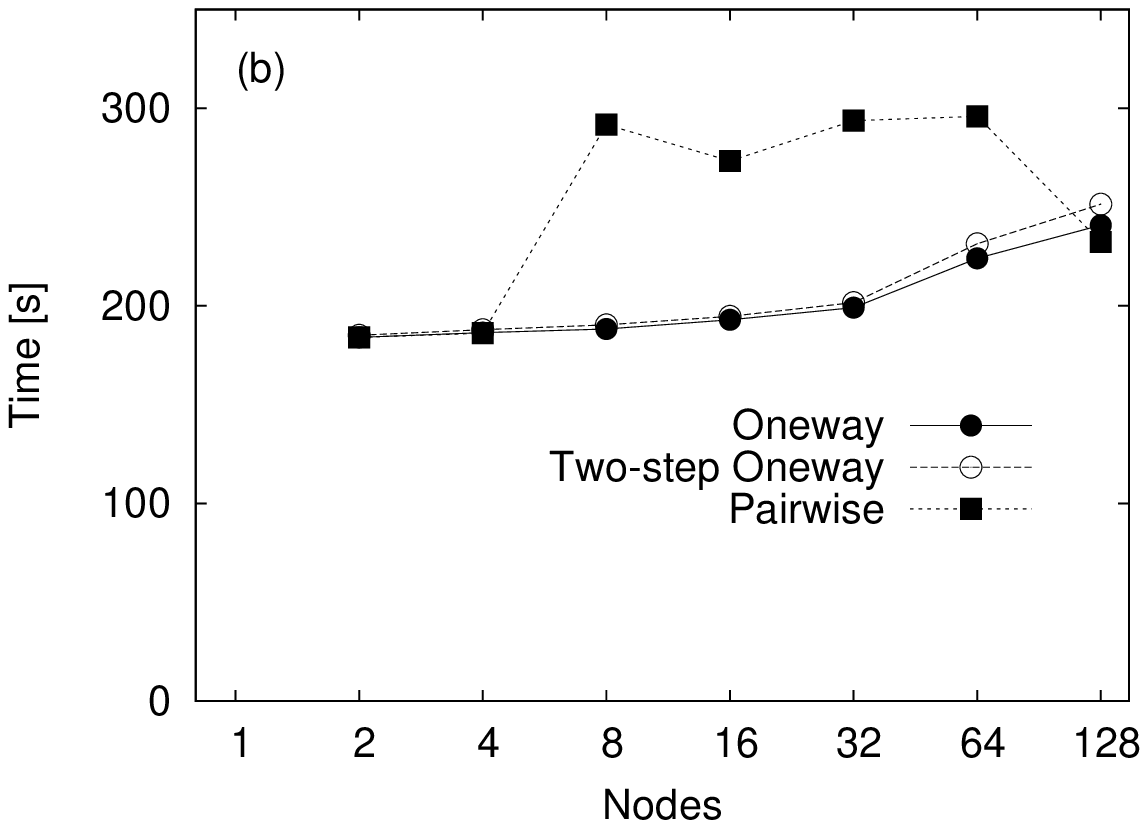}
\end{center}
\caption{ Comparison between three communication schemes at (a) NIFS and (b) ISSP.
The Oneway scheme exhibits the best performance at both facilities.
}
\label{fig_pairwise_results}
\end{figure}

A comparison of benchmark test for above the three methods is shown in Fig.~\ref{fig_pairwise_results}.
Figure~\ref{fig_pairwise_results}~(a) shows the results at NIFS.
The results of Pairwise only slightly differ from those of Oneway.
While the results of Two-step Oneway show similar performance up to 32 nodes,
those for larger nodes are significantly slower than the other methods.
Figure~\ref{fig_pairwise_results}~(b) shows the results at ISSP.
There are major differences between the performances of the three methods even though the amounts of communication are identical, and Two-step Oneway considerably slower than the other methods.
The reason why Two-step Oneway is the slowest method in both cases may be that 
double threads are executed by separating the sending and receiving processes instead of using MPI\_Sendrecv.
While it is unclear why the results of the three methods are almost the same for the largest number of nodes at ISSP,
we can conclude that the overhead of communications strongly depends on the hardware environment. 
Considering the above results at NIFS and ISSP,  the simple Oneway scheme is found to be less affected by
the architecture of parallel machine and consistently has relatively good performance.

\subsection{Synchronicity Problem} \label{sec_osjitter}

Parallel programs adopting the single program multiple data (SPMD) model usually require synchronicity,
and the parallel MD described here is one such example.
The slowest process determines the overall performance,
since other processes must wait for the slowest process owing to the synchronization.
Therefore, the parallel efficiency is strongly affected by the fluctuation of the computation time of each process.
One of the factors causing the fluctuation is load imbalance.
Load imbalance refers to the inhomogeneity of workload among processes.
Unequally assigned tasks cause the fluctuation of the computational time of each step
and they decrease the parallel efficiency.
The workload of parallel MD mainly depends on the number of interacting pairs in each process.
To clarify this problem quantitatively, the number of interacting pairs is counted 
at the final state of the 128-node run.
The parallel efficiency is determined by the slowest process which should be with the largest number of interacting pairs.

In our calculation, we found that the largest and the smallest number of
pairs were $11,656,318$ and $11,629,010$, respectively.
The difference is about $0.02\%$, which cannot be responsible for the reduction in parallel efficiency.
Note that this small fluctuation originates from the fact that the initial condition is homogeneous.
The load imbalance can be serious for inhomogeneous systems such as systems involving phase transitions.
However, our benchmark results show that the parallel efficiency decreases as the number of processes increases,
even for a homogeneous system with a uniform workload for each process.
Therefore, we assumed that there are other factors that affect the parallel efficiency.

\begin{figure}[tb]
\begin{center}
\includegraphics[width=0.49\linewidth]{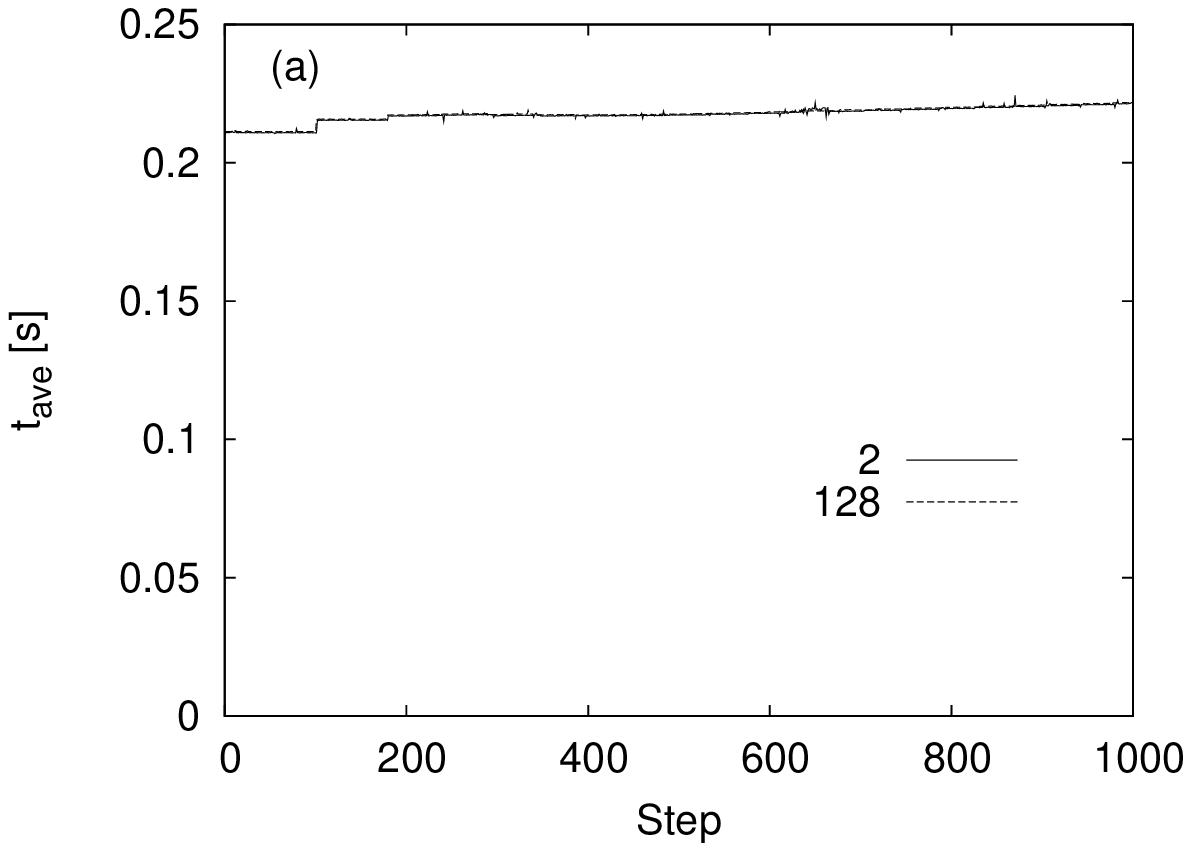}
\includegraphics[width=0.49\linewidth]{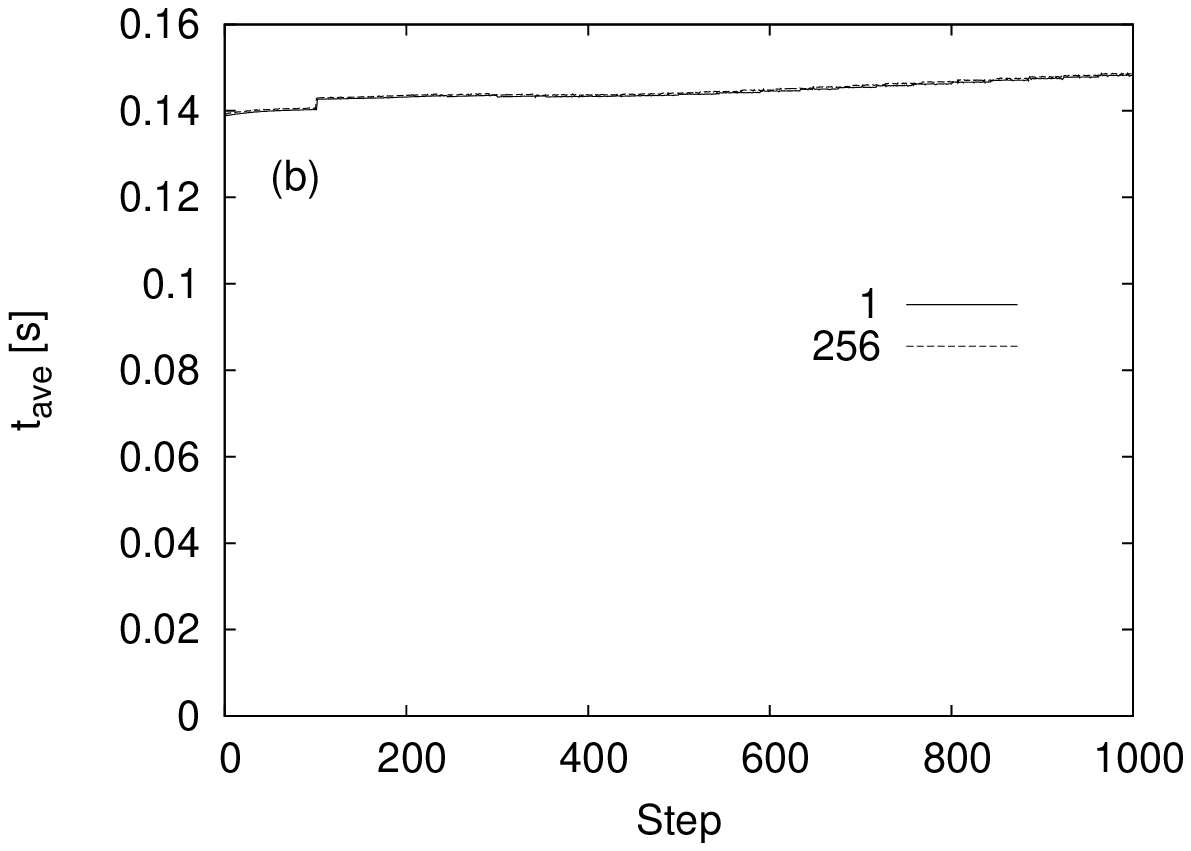}
\end{center}
\caption{
Computational CPU time for force calculations averaged over all processes at each step.
(a) The data obtained at NIFS. The data obtained from the smallest and largest runs are denoted by `2' and `128', respectively. 
(b) The data obtained at ISSP. The data obtained from the smallest and largest runs are denoted by `1' and `256', respectively. 
Time evolutions are almost independent of the number of nodes.
}
\label{fig_ave}
\end{figure}

\begin{figure}[tb]
\begin{center}
\includegraphics[width=0.49\linewidth]{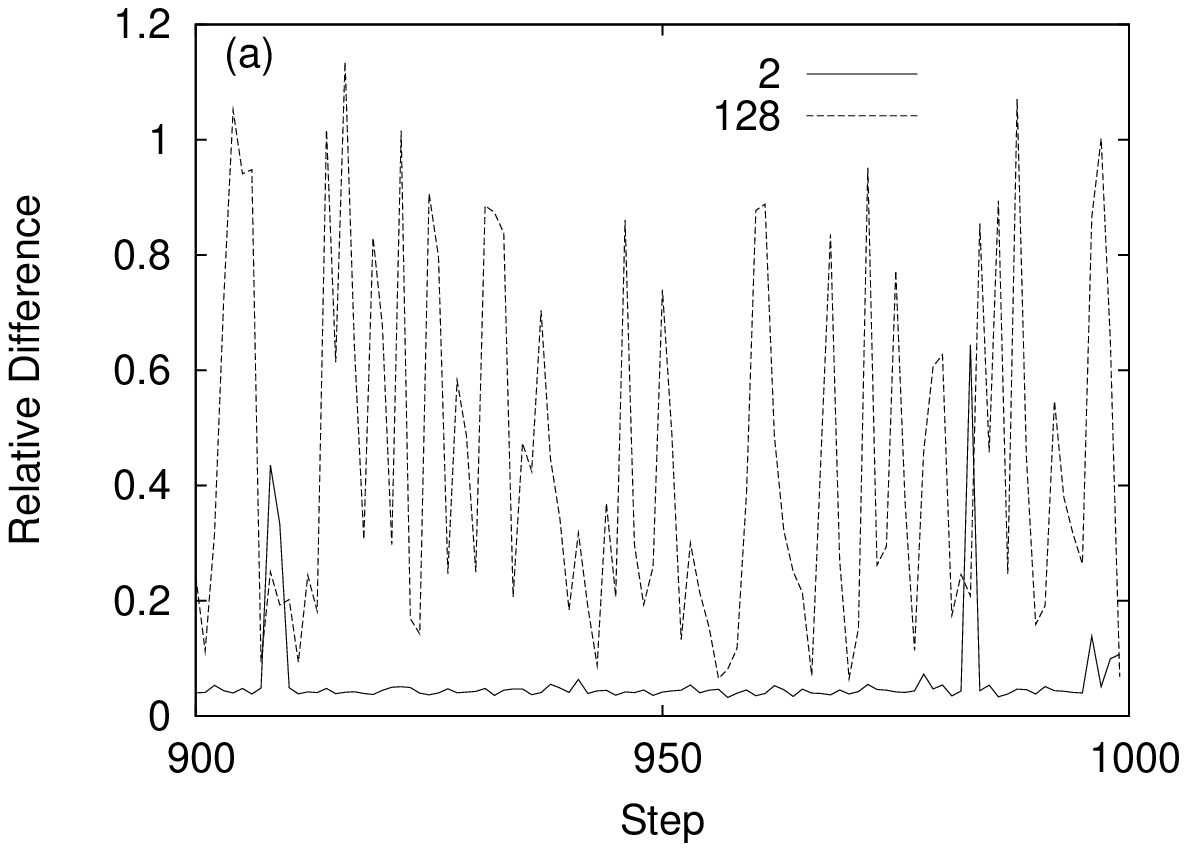}
\includegraphics[width=0.49\linewidth]{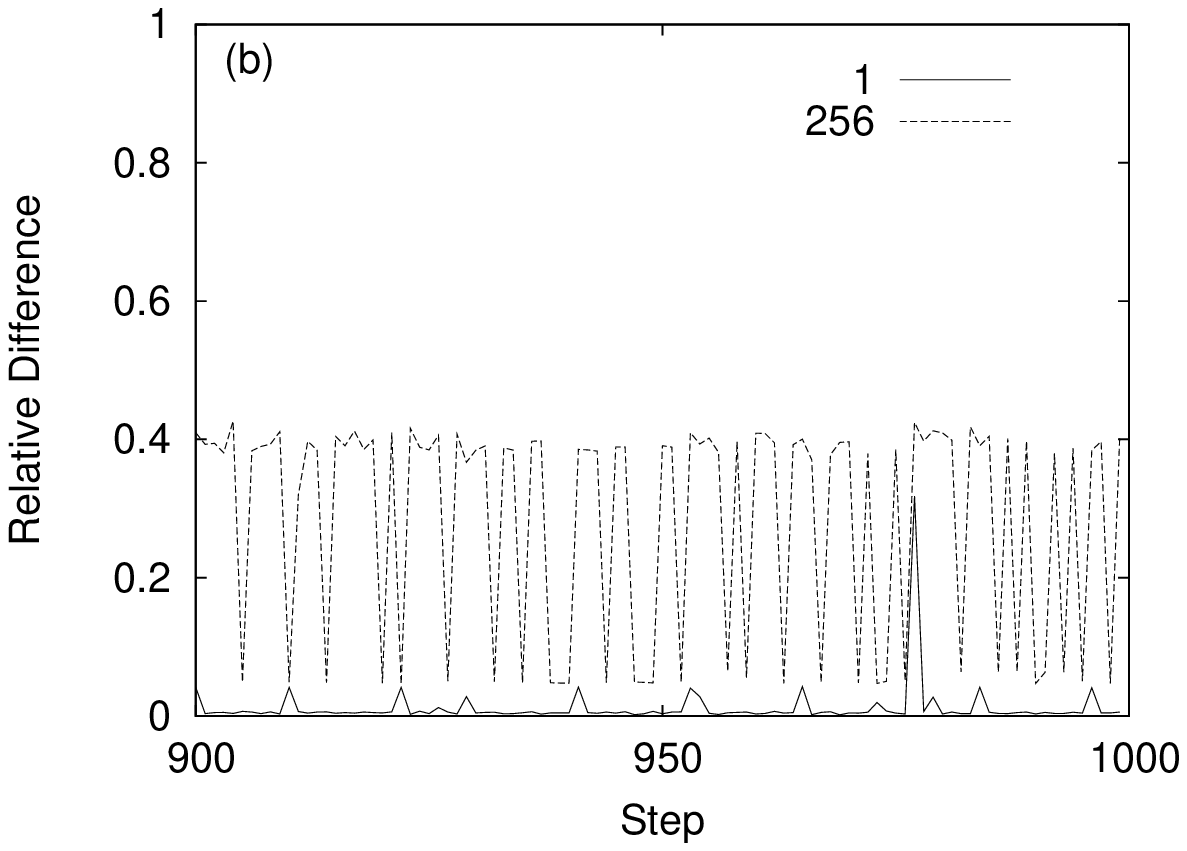}
\end{center}
\caption{
Time evolutions of the relative difference at (a) NIFS and (b) ISSP.
Only the final 100 steps are shown for the visibility.
The fluctuations becomes larger as the number of nodes increases.
}
\label{fig_fluc}
\end{figure}

Next, we investigate the fluctuation of the computational time of force for each process.
Suppose $t^{(i)}_{(p)}$ is the computational time spent for force calculation of $p$-th process at $i$-th step.
The average of the force calculation at $i$-th step is defined as
\begin{equation}
t_\mathrm{ave}^{(i)} = \frac{1}{N_p} \sum_p^{N_p} t^{(i)}_{(p)},
\end{equation}
with the number of processes $N_p$.
The time evolutions of $t_\mathrm{ave}^{(i)}$ at NIFS and ISSP are shown in Fig.~\ref{fig_ave}.
The data of the smallest and the largest runs are presented.
Note that, we observe the time spent only for force calculation which
dose not involve communication.
We found that the computational time for force
show almost same values and independent of the number of nodes.
However, the compuational time for each step itself might fluctuate,
which cause the one of candidate of reduction of parallel efficiency.
Therefore, we focus on the calculation time of force for each step.
To investigate the fluctuation of the computational time for each step,
we define the relative difference as
\begin{equation}
t_\mathrm{diff}^{(i)} = \frac{ t_\mathrm{max}^{(i)} - t_\mathrm{min}^{(i)} }{ t_\mathrm{ave}^{(i)}},
\end{equation}
where $t_\mathrm{max}^{(i)} \equiv \max_p \{t_{(p)}^{(i)}  \}$ is the maximum time, and
$t_\mathrm{min}^{(i)} \equiv \min_p \{t_{(p)}^{(i)}  \}$ is the minimum time spent on the force calculation among 
all processes at $i$th step~\cite{fluctuation}.
$t_\mathrm{diff}^{(i)}$ denotes the difference between the times spent on the slowest and fastest processes normalized by the average time at $i$th step.
The time evolutions of the average and relative difference for force calculations at NIFS are shown in Fig.~\ref{fig_fluc}~(a).
While the average times for force calculations are identical for 2- and 128-node runs,
the relative differences of the large run are significantly larger than those of the small run.
The relative difference of the large run sometimes fluctuates by 100\% ($t_\mathrm{diff}^{(i)}$ becomes 1.0),
which implies that the computational time is doubled at the step.
The same plots for ISSP are shown in Fig.~\ref{fig_fluc}~(b).
Similar phenomena are observed, although the relative differences of the large run are at most 40\%.
The relative differences of the small run at ISSP are almost zero, since the run has only four processes,
which is the reason why the fluctuation is relatively small.

From the above, we can conclude that the main factor degrading the parallel efficiency is
the fluctuation of the computational time at each step.
Although the average time varies only slightly,
the fluctuation of the computational time between processes increases as the number of processes increases.
Then the delays of the slowest process accumulate as the overhead.
There are several possible factors causing this fluctuation.
One possible source is noise from an operating system (OS).
An OS has several background processes called daemons, which interfere with the user execution time and decreases its efficiency~\cite{Beckman2006}.
This noise from OS is referred to as OS jitter or OS detour.
Interruptions by an operating system occur randomly, and they cause the fluctuation of the execution time of each process.
Not only interruptions themselves, but also side effects caused by the interruptions, such as pollution of the CPU cache, are also a source of noise.

\section{Summary and Further Issues}
\label{sec_summary}

We have developed an MD code that is not only suitable for massively parallel processors
but also exhibits good single-processor performance for IBM Power and Intel Xeon architectures.
Most of all algorithms presented in this paper can be applied to other particle systems
with short-range interactions including many-body interactions.
There are three levels of optimization. At the highest level, we attempt to reduce the 
computational cost by developing or choosing appropriate algorithms such as the bookkeeping method.
Next, we have to manage memory access. All data required by cores should be in the cache
so that the cores make the best use of its computational power.
Finally, we have to perform architecture-dependent tuning, since
the development paradigms are considerably different between different CPUs.

MD is suitable for MPP-type simulations since the computation-communication ratio is
generally small. However, we have observed a significant degradation in parallel efficiency
despite the use of conditions under which the communication cost should be negligible.
We have found that the fluctuation of the execution time of processes reduces the parallel efficiency.
The fluctuations become serious when the number of process becomes over one thousand.
The OS jitter, which is the interference by the operating system, can be a source of the fluctuation.
There are many attempts to handle the OS jitter, but it is difficult to be settled by user programs.
One possibility is to adopt hybrid parallelization, \textit{i.e.},
use OpenMP for inner-node communications and use MPI for internode communications.
Hybrid parallelization not only reduces the number of total MPI processes,
but also averages out the fluctuations which can improve the parallel efficiency.
Note that the affect due to the OS jitter is generally order of nanoseconds or shorter,
and the time required to perform one step of MD simulations is order of millisecond.
Therefore, there may be other sources of noise, such as translation lookaside buffer (TLB) miss.
TLB is a kind of cache, which translate the address from virtual to real~\cite{Tanenbaum}.
While it makes computation faster, it takes long time to fetch the data 
if the required address is not cached in the buffer. The translation tables are somtimes
swapped out into other storage such as hard-disk, then the penalty of TLB miss can be expensive.
Investigation of the source of the fluctuation in execution time is one of the further issues.

In this paper, the load imbalance problem is not considered.
Load balancing is important for parallelization with a huge number of processes since
the fluctuation of the timing of each process greatly degrades the parallel efficiency
as shown in Sec.~\ref{sec_osjitter}.
There are two approaches to load balancing, one is based on
force decomposition and the other is based on domain decomposition.
The load balancing in NAMD (Not Another Molecular Dynamics program)~\cite{NAMD} is essentially achieved by the force decomposition,
while NAMD adopts a hybrid strategy of parallelization.
The processes of NAMD are under the control of a dynamic load balancer.
The simulation box is divided into subdomains (called \textit{patches}), and
the total computational task is divided into subtasks (called \textit{compute objects}).
The load balancer measures the load on each process and reassigns \textit{compute objects}
to processors if necessary.
GROMACS (Groningen Machine for Chemical Simulations)~\cite{GROMACS} adopts a domain-decomposition-based strategy for load balancing,
that is, it changes the volume of domains assigned to processors to
improve the load balance. The simulation box is divided into staggered grids
whose volume are different. A load balancer changes the volume of each grid
by moving the boundaries of cells to reduce load imbalance.
There is also a method of non-box-type space decomposition that utilizes the
Voronoi construction. Reference points are placed in a simulation box and 
each Voronoi cell is assigned to a process. Load balancing is performed
by changing the positions of the reference points~\cite{Yahagi1999,Koradi2000}.
Generally speaking, force-decomposition-based load balancing is a better choice 
if a system contains long-range interactions such as electrostatic force,
and domain-decomposition-based load balancing is better for a system with short-range interactions.
However, the choice strongly depends on the phenomenon to be simulated, and therefore, 
both strategies or some hybrid of them should be considered.

The source codes used in this paper has been published online~\cite{source}.
We hope that the present paper and our source codes will help researchers
to develop their own parallel MD codes that can utilize the computational power
of petaflop, and eventually exaflop machines.

\section*{Acknowledgements}
The authors would like to thank Y. Kanada, S. Takagi, and T. Boku for fruitful discussions.
Some parts of the implementation techniques are owing to N. Soda and M. Itakura.
HW thanks M. Isobe for useful information of past studies.
This work was supported by KAUST GRP (KUK-I1-005-04), Grants-in-Aid for Scientific Research (Contracts No.\ 19740235),
and the NIFS Collaboration Research program (NIFS10KTBS006).
The computations were carried out using the facilities of National Institute for Fusion Science;
the Information Technology Center, the University of Tokyo; 
the Supercomputer Center, Institute for Solid State Physics, University of Tokyo;
and the Research Institute for Information Technology, Kyushu University.

%


\begin{thebibliography}{99}
\bibitem{Alder1957} B. J. Alder and T. E. Wainwright, J. Chem. Phys. \textbf{27}  (1957), 1208.

\bibitem{Rahman1964} A. Rahman, \PR{136,1964,405}. 

\bibitem{Allen} M. P. Allen and D. J. Tildesley, Computer Simulation of Liquids, (Clarendon Press, Oxford 1987).

\bibitem{Frenkel} D. Frenkel and B. Smit, Understanding Molecular Simulation -- From Algorithms to Applications, (Academic Press, NewYork, 2001)

\bibitem{Rapaport} D. Rapaport, The art of molecular dynamics simulation, (Cambridge University Press, 2004).

\bibitem{HooverBook} W. G. Hoover, Molecular Dynamics, Lecture Notes in Physics \textbf{258} (Springer-Verlag, Berlin, 1986).

\bibitem{NAMD} L. Kal\'e, R. Skeel, M. Bhandarkar, R. Brunner, A. Gursoy, N. Krawetz, J. Phillips, A. Shinozaki, K. Varadarajan, and K. Schulten, J. Comput. Phys. \textbf{151}, (1999), 283.

\bibitem{GROMACS} B. Hess, C. Kutzner, D. van der Spoel, and E. Lindahl, J. Chem. Theory Comput. \textbf{4} (2008) 435.

\bibitem{Car1985} R. Car and M. Parrinello, \PRL{55,1985,2471}.

\bibitem{Top500} \verb|http://www.top500.org/|

\bibitem{RIKEN} \verb|http://www.nsc.riken.jp/|

\bibitem{Plimpton1995} S. Plimpton, J. Comput. Phys. \textbf{117} (1995), 117.

\bibitem{Germann2008} T. C. Germann and K. Kadau, \IJMP{C19,2008,1315}.

\bibitem{Holian1991}  B. L. Holian, A. F. Voter, N. J. Wagner, R. J. Ravelo, and S. P. Chen, \PRA{43,1991,2655}.

\bibitem{Beazley1994} D. M. Beazley and P. S. Lomdahl, Parallel Comput. \textbf{20} (1994), 173.

\bibitem{Spotswood1973} S. D. Stoddard and J. Ford, \PRA{8,1973,1504}.

\bibitem{Quentrec1975} B. Quentrec and C. Brot, J. Comput. Phys. \textbf{13} (1975), 430.

\bibitem{Hockney1981} R. W. Hockney and J. W. Eastwood, Computer Simulation Using Particles (McGraw-Hill, New York, 1981).

\bibitem{Form1993} W. Form, N, Ito, and G. A. Kohring, \IJMP{C4,1993,1085}.

\bibitem{Knuth1973} D. E. Knuth, The Art of Computer Programming (Addison--Wesley, Reading, MA, 1973).

\bibitem{Grest1989} G. S. Grest and B. D\"unweg, Comput. Phys. Commun. \textbf{55} (1989) 269.

\bibitem{Kadau2006} K. Kadau, T. C. Germann, and P. S. Lomdahl, \IJMP{C17,2006,1755}.

\bibitem{Verlet1967} L. Verlet, \PR{195,1967,98}.

\bibitem{Isobe1999} M. Isobe, \IJMP{C10,1999,1281}.

\bibitem{Meloni2007} S. Meloni, M. Rosati, and L. Colombo, J. Chem. Phys. \textbf{126} (2007), 121102.

\bibitem{Hayashi} R. Hayashi, K. Tanaka, S. Horiguchi, and Y. Hiwatari, Parallel Processing Letters, \textbf{15},  (2005) 481.

\bibitem{Beazley} D.M. Beazley, P.S. Lomdahl, N.G. Jensen, R. Giles,  and P. Tamayo, Parallel Algorithms for Short-Range Molecular Dynamics.,  World Scientific's Annual Reviews in Computational Physics, \textbf{3}, (1995) 119. 

\bibitem{MaximumVelocity} In equilibrium, a distribution for the velocity obeys the Maxwell--Boltzmann type.
While there can be an infinitely fast particle in an infinite-size system,
the exectation value of the maximum velocity is finite in the finite-size system
and the value increases as the system size increases.

\bibitem{fluctuation} We investigated the difference between the maximum and the minimum time spent for 
force calculation amoung processes instead of the standard deviation, since the slowest process determine
the global speed of the calculation and the quantity observed in the present manuscript
is directly associated with the parallel overhead.
It is also worth to be noted that the probability distribution function of $t^{(i)}_{(p)}$ is not 
the Gaussian type. Therefore, the standard deviation is not an appropriate index to characterize the
fluctuation of the calculation time.

\bibitem{Beckman2006} P. Beckman, K. Iskra, K. Yoshii, and S. Coghlan, ACM SIGOPS Operating Systems Review \textbf{40}, (2006) 29.

\bibitem{Tanenbaum} A. S. Tanenbaum, Modern Operating Systems, (Prentice Hall, 2001).

\bibitem{Yahagi1999} H. Yahagi, M. Mori, and Y. Yoshii, ApJS, \textbf{124} (1999), 1.

\bibitem{Koradi2000} R. Koradi, M. Billeter, and P. Guinter, Comput. Phys. Comm. \textbf{124} (2000), 139.

\bibitem{source} \verb|http://mdacp.sourceforge.net/|


\end{thebibliography}
\end{document}